\documentclass[prd,aps,twocolumn,amsfonts,showpacs,superscriptaddress,preprintnumbers,nofootinbib,a4paper]{revtex4-2}

\usepackage{graphicx}
\usepackage{xcolor}
\usepackage{rotating}
\usepackage{bm,amsmath,amssymb}
\usepackage[utf8]{inputenc}
\usepackage{footmisc}
\usepackage{diagbox}
\usepackage{lipsum, babel}
\usepackage{cancel}

\usepackage[normalem]{ulem}
\usepackage{tikz}
\usepackage{longtable}
\usepackage{graphicx}
\usepackage{amsmath}
\usepackage{amssymb}
\usepackage{mathtools}
\usepackage{xstring}
\usetikzlibrary{arrows.meta}
\usetikzlibrary{decorations.markings}
\usetikzlibrary{snakes}
\usetikzlibrary{shapes.geometric}
\usetikzlibrary{calc}

\usepackage[colorlinks=true, urlcolor=blue, linkcolor=blue, citecolor=blue, hyperindex=true, linktocpage=true]{hyperref}

\newcommand{\nn}{\nonumber\\}

\def\im{\mbox{Im}}

\def\le{\left}
\def\ri{\right}

\newcommand{\dd}{\mathrm{d}}

\def\be{\begin{equation}}
\def\ee{\end{equation}}

\def\bea{\begin{eqnarray}}
\def\eea{ \end{eqnarray}}
\newcommand{\bma}{\le(\begin{matrix}}
\newcommand{\ema}{\end{matrix}\ri)}

\newcommand{\bega}{\begin{gather}}

\def\im{\text{Im}}

\tikzset{snek/.style={decorate, decoration={snake, segment length=1.5mm, amplitude=.5mm}}}
\tikzset{snekL/.style={decorate, decoration={snake, segment length=1.5mm, amplitude=-.5mm}}}

\def\centerarc[#1](#2)(#3:#4:#5) { \draw[#1] ($(#2)+({#5*cos(#3)},{#5*sin(#3)})$) arc (#3:#4:#5); }

\def\typeAnum[#1, #2, #3, #4, #5, #6, #7, #8, #9] {
    \begin{tikzpicture}[baseline]
    \begin{scope}[very thick]
    \draw[] (-.75 - 2 * .75 *pi/4,0)--(-.75 - .75*pi/4,0);
    \draw[snekL] (-.75 - .75*pi/4,0)--(-.75,0);
    \draw[] (0,.75)--(0,0);
    \draw[#9] (0,0)--(0,-.75);
    \draw[] (.75 + 2 * .75*pi/4,0)--(.75 + .75*pi/4,0);
    \draw[snek] (.75 + .75*pi/4,0)--(0.75,0);
    \centerarc[#1](0,0)(-90:-45:.75)
    \centerarc[#2](0,0)(-45:0:.75)
    \centerarc[#3](0,0)(0:45:.75)
    \centerarc[#4](0,0)(45:90:.75)
    \centerarc[#8](0,0)(135:90:.75)
    \centerarc[#7](0,0)(180:135:.75)
    \centerarc[#6](0,0)(225:180:.75)
    \centerarc[#5](0,0)(270:225:.75)
    \end{scope}
    \end{tikzpicture}
}

\def\typeAres[#1, #2, #3, #4, #5, #6, #7, #8, #9] {
    \begin{tikzpicture}[baseline]
    \begin{scope}[very thick]
    \draw[] (-.75 - 2 * .75 *pi/4,0)--(-.75 - .75*pi/4,0);
    \draw[snekL] (-.75 - .75*pi/4,0)--(-.75,0);
    \draw[] (0,.75)--(0,0);
    \draw[#9] (0,0)--(0,-.75);
    \draw[] (.75 + 2 * .75*pi/4,0)--(0.75,0);
    \centerarc[#1](0,0)(-90:-45:.75)
    \centerarc[#2](0,0)(-45:0:.75)
    \centerarc[#3](0,0)(0:45:.75)
    \centerarc[#4](0,0)(45:90:.75)
    \centerarc[#8](0,0)(135:90:.75)
    \centerarc[#7](0,0)(180:135:.75)
    \centerarc[#6](0,0)(225:180:.75)
    \centerarc[#5](0,0)(270:225:.75)
    \end{scope}
    \end{tikzpicture}
}

\def\typeBnum[#1] {
    \begin{tikzpicture}[baseline={([yshift=-.8ex]current bounding box.center)}]
    \begin{scope}[very thick]
    \draw[] (-1.5,0)--(-.75,0);
    \draw[snek] (-.75,0)--(0,0);
    \draw[snek] (0,0)--(.75,0);
    \draw[] (.75,0)--(1.5,0);
    \centerarc[](0,.75)(180:360:.75)
    \centerarc[snekL](0,.75)(180:90:.75)
    \centerarc[#1](0,.75)(0:90:.75)
    \centerarc[](0,1.125)(0:360:.375)
    \end{scope}
    \end{tikzpicture}
}

\def\typeBres[#1] {
    \begin{tikzpicture}[baseline={([yshift=-.8ex]current bounding box.center)}]
    \begin{scope}[very thick]
    \draw[] (-1.5,0)--(-.75,0);
    \draw[snek] (-.75,0)--(0,0);
    \draw[] (0,0)--(1.5,0);
    \centerarc[](0,.75)(180:360:.75)
    \centerarc[snekL](0,.75)(180:90:.75)
    \centerarc[#1](0,.75)(0:90:.75)
    \centerarc[](0,1.125)(0:360:.375)
    \end{scope}
    \end{tikzpicture}
}

\def\typeCnum[#1, #2] {
    \begin{tikzpicture}[baseline]
    \begin{scope}[very thick]
    \draw[] (-.75 - 2 * .75 *pi/4,0)--(-.75 - .75*pi/4,0);
    \draw[snek] (-.75 - .75*pi/4,0)--(-.75,0);
    \draw[] (0,0)--(.75,0);
    \draw[#1] (0,0)--(-.75,0);
    \draw[] (.75 + 2 * .75*pi/4,0)--(.75 + .75*pi/4,0);
    \draw[snek] (.75 + .75*pi/4,0)--(0.75,0);
    \centerarc[](0,0)(0:270:.75)
    \centerarc[#2](0,0)(270:360:.75)
    \end{scope}
    \end{tikzpicture}
}

\def\typeCres[#1] {
    \begin{tikzpicture}[baseline]
    \begin{scope}[very thick]
    \draw[] (-.75 - 2 * .75 *pi/4,0)--(-.75 - .75*pi/4,0);
    \draw[snekL] (-.75 - .75*pi/4,0)--(-.75,0);
    \draw[snek] (0,0)--(.75,0);
    \draw[] (0,0)--(-.75,0);
    \draw[] (.75 + 2 * .75*pi/4,0)--(0.75,0);
    \centerarc[](0,0)(0:270:.75)
    \centerarc[#1](0,0)(270:360:.75)
    \end{scope}
    \end{tikzpicture}
}

\def\typeDnum[#1, #2, #3, #4] {
    \begin{tikzpicture}[baseline={([yshift=-.8ex]current bounding box.center)}]
    \begin{scope}[very thick]
    \draw[] (-1.25,0)--(-.75,0);
    \draw[snekL] (-.75,0)--(0,0);
    \draw[#1] (0,0)--(.75,0);
    \draw[#2] (1.5,0)--(.75,0);
    \draw[] (0,0)--(.375,.375);
    \draw[snekL] (.375,.375)--(.75,.75);
    \draw[#4] (1.125,.375)--(.75,.75);
    \draw[#3] (1.125,.375)--(1.5,0);
    \draw[snek] (2.25,0)--(1.5,0);
    \draw[] (2.25,0)--(2.75,0);
    \centerarc[](.75,1.125)(0:360:.375)
    \end{scope}
    \end{tikzpicture}
}

\def\typeDres[#1, #2, #3, #4] {
    \begin{tikzpicture}[baseline={([yshift=-.8ex]current bounding box.center)}]
    \begin{scope}[very thick]
    \draw[] (-1.25,0)--(-.75,0);
    \draw[snekL] (-.75,0)--(0,0);
    \draw[] (0,0)--(.75,0);
    \draw[#1] (1.5,0)--(.75,0);
    \draw[] (0,0)--(.375,.375);
    \draw[#4] (.375,.375)--(.75,.75);
    \draw[#3] (1.125,.375)--(.75,.75);
    \draw[#2] (1.125,.375)--(1.5,0);
    \draw[] (1.5,0)--(2.75,0);
    \centerarc[](.75,1.125)(0:360:.375)
    \end{scope}
    \end{tikzpicture}
}

\def\typeEnum[#1, #2] {
    \begin{tikzpicture}[baseline={([yshift=-.8ex]current bounding box.center)}]
    \begin{scope}[very thick]
    \draw[] (-2,0)--(-1.6875,0);
    \draw[snekL] (-1.6875,0)--(-1.125,0);
    \draw[] (2,0)--(1.6875,0);
    \draw[snek] (1.6875,0)--(1.125,0);
    \centerarc[](-0.5625,0)(360:90:0.5625)
    \centerarc[snekL](-0.5625,0)(90:0:0.5625)
    \centerarc[](0.5625,0)(180:360:0.5625)
    \centerarc[#2](0.5625,0)(0:90:0.5625)
    \centerarc[#1](0.5625,0)(90:180:0.5625)
    \end{scope}
    \end{tikzpicture}
}

\def\typeEres[#1, #2] {
    \begin{tikzpicture}[baseline={([yshift=-.8ex]current bounding box.center)}]
    \begin{scope}[very thick]
    \draw[] (-2,0)--(-1.6875,0);
    \draw[snekL] (-1.6875,0)--(-1.125,0);
    \draw[] (2,0)--(1.125,0);
    \centerarc[#1](-0.5625,0)(360:270:0.5625)
    \centerarc[](-0.5625,0)(270:90:0.5625)
    \centerarc[snekL](-0.5625,0)(90:0:0.5625)
    \centerarc[](0.5625,0)(270:90:0.5625)
    \centerarc[snek](0.5625,0)(0:90:0.5625)
    \centerarc[#2](0.5625,0)(270:360:0.5625)
    \end{scope}
    \end{tikzpicture}
}

\def\typeFnum[#1, #2, #3, #4]{
    \begin{tikzpicture}[baseline={([yshift=-.8ex]current bounding box.center)}]
    \begin{scope}[very thick]
    \draw[] (-1.5,0)--(-.75,0);
    \draw[snekL] (-.75,0)--(0,0);
    \draw[snek] (.75,0)--(0,0);
    \draw[] (1.5,0)--(.75,0);
    \centerarc[#1](0,0)(60:90:1.5)
    \centerarc[](0,0)(90:120:1.5)
    \centerarc[](-.75,{.75 * sqrt(3)})(300:330:1.5)
    \centerarc[#4](-.75,{.75 * sqrt(3)})(330:360:1.5)
    \centerarc[snekL](.75,{.75 * sqrt(3)})(210:180:1.5)
    \centerarc[](.75,{.75 * sqrt(3)})(210:240:1.5)
    \centerarc[#2](0,{1.5 * sqrt(3)})(240:270:1.5)
    \centerarc[#3](0,{1.5 * sqrt(3)})(270:300:1.5)
    \end{scope}
    \end{tikzpicture}
}

\def\typeFres[#1, #2, #3, #4]{
    \begin{tikzpicture}[baseline={([yshift=-.8ex]current bounding box.center)}]
    \begin{scope}[very thick]
    \draw[] (-1.5,0)--(-.75,0);
    \draw[snekL] (-.75,0)--(0,0);
    \draw[] (1.5,0)--(0,0);
    \centerarc[#1](0,0)(60:90:1.5)
    \centerarc[](0,0)(90:120:1.5)
    \centerarc[](-.75,{.75 * sqrt(3)})(300:330:1.5)
    \centerarc[#4](-.75,{.75 * sqrt(3)})(330:360:1.5)
    \centerarc[snekL](.75,{.75 * sqrt(3)})(210:180:1.5)
    \centerarc[](.75,{.75 * sqrt(3)})(210:240:1.5)
    \centerarc[#2](0,{1.5 * sqrt(3)})(240:270:1.5)
    \centerarc[#3](0,{1.5 * sqrt(3)})(270:300:1.5)
    \end{scope}
    \end{tikzpicture}
}

\def\typeGnum[#1, #2, #3, #4, #5, #6, #7]{
    \begin{tikzpicture}[baseline={([yshift=-.8ex]current bounding box.center)}]
    \begin{scope}[very thick]
    \draw[] (-1.25, 0)--(-.75, 0);
    \draw[snek] (-.75, 0)--(0, 0);
    \draw[snek] (1.5, 0)--(2.25, 0);
    \draw[] (2.25, 0)--(2.75, 0);
    \centerarc[#7](0, 0)(0:30:1.5)
    \centerarc[#6](0, 0)(30:60:1.5)
    \centerarc[#1](.75, {sqrt(3) * .75})(240:270:1.5)
    \centerarc[#2](.75, {sqrt(3) * .75})(270:300:1.5)
    \centerarc[#3](1.5, 0)(180:150:1.5)
    \centerarc[#4](1.5, 0)(150:120:1.5)
    \centerarc[#5](2.25, {sqrt(3) * .75})(180:210:1.5)
    \centerarc[](2.25, {sqrt(3) * .75})(210:240:1.5)
    \end{scope}
    \end{tikzpicture}
}

\def\typeGres[#1, #2, #3, #4, #5, #6, #7]{
    \begin{tikzpicture}[baseline={([yshift=-.8ex]current bounding box.center)}]
    \begin{scope}[very thick]
    \draw[] (-1.25, 0)--(-.75, 0);
    \draw[snek] (-.75, 0)--(0, 0);
    \draw[] (1.5, 0)--(2.75, 0);
    \centerarc[#7](0, 0)(0:30:1.5)
    \centerarc[#6](0, 0)(30:60:1.5)
    \centerarc[#1](.75, {sqrt(3) * .75})(240:270:1.5)
    \centerarc[#2](.75, {sqrt(3) * .75})(270:300:1.5)
    \centerarc[#3](1.5, 0)(180:150:1.5)
    \centerarc[#4](1.5, 0)(150:120:1.5)
    \centerarc[#5](2.25, {sqrt(3) * .75})(180:210:1.5)
    \centerarc[](2.25, {sqrt(3) * .75})(210:240:1.5)
    \end{scope}
    \end{tikzpicture}
}

\def\typeGresalt[#1, #2, #3, #4, #5, #6]{
    \begin{tikzpicture}[baseline={([yshift=-.8ex]current bounding box.center)}]
    \begin{scope}[very thick]
    \draw[] (-1.25, 0)--(0, 0);
    \draw[snek] (1.5, 0)--(2.25, 0);
    \draw[] (2.25, 0)--(2.75, 0);
    \centerarc[#6](0, 0)(0:30:1.5)
    \centerarc[#5](0, 0)(30:60:1.5)
    \centerarc[#1](.75, {sqrt(3) * .75})(240:270:1.5)
    \centerarc[](.75, {sqrt(3) * .75})(270:300:1.5)
    \centerarc[#2](1.5, 0)(180:150:1.5)
    \centerarc[#3](1.5, 0)(150:120:1.5)
    \centerarc[#4](2.25, {sqrt(3) * .75})(180:210:1.5)
    \centerarc[](2.25, {sqrt(3) * .75})(210:240:1.5)
    \end{scope}
    \end{tikzpicture}
}

\def\gee(#1){G_{\varepsilon\varepsilon}(#1)}
\def\gfe(#1){G_{\varphi\varepsilon}(#1)}
\def\gef(#1){G_{\varepsilon\varphi}(#1)}

\begin{document}

%\preprint{APS/123-QED}
\title{
 Analytic structure of diffusive correlation functions 
}

\author{Sa\v{s}o Grozdanov}
\affiliation{Higgs Centre for Theoretical Physics, University of Edinburgh, Edinburgh, EH8 9YL, Scotland}
\affiliation{Faculty of Mathematics and Physics, University of Ljubljana, Jadranska ulica 19, SI-1000 Ljubljana, Slovenia}
 
\author{Timotej Lemut}
 \affiliation{Faculty of Mathematics and Physics, University of Ljubljana, Jadranska ulica 19, SI-1000 Ljubljana, Slovenia}

\author{Jaka Pelai\v{c}}
\affiliation{Mathematical Institute, University of Oxford, 	Woodstock Road, OX2 6GG, England
}

\author{Alexander Soloviev}
\affiliation{Faculty of Mathematics and Physics, University of Ljubljana, Jadranska ulica 19, SI-1000 Ljubljana, Slovenia}

\begin{abstract}
Diffusion is a dissipative transport phenomenon ubiquitously present in nature. Its details can now be analysed with modern effective field theory (EFT) techniques that use the closed-time-path (or Schwinger-Keldysh) formalism. We discuss the structure of the diffusive effective action appropriate for the analysis of stochastic or thermal loop effects, responsible for the so-called long-time tails, to all orders. We also elucidate and prove a number of properties of the EFT and use the theory to establish the analytic structure of the $n$-loop contributions to diffusive retarded two-point functions. Our analysis confirms a previously proposed result by Delacr\'{e}taz that used microscopic conformal field theory arguments. Then, we analyse a number of implications of these loop corrections to the dispersion relations of the diffusive mode and new, gapped modes that appear when the EFT is treated as exact. Finally, we discuss certain features of an all-loop model of diffusion that only retains a special subset of $n$-loop `banana' diagrams.
\end{abstract}

\maketitle

%\tableofcontents
\section{Introduction}

Hydrodynamic transport phenomena underpin much of theoretical and experimental research. Such phenomena arise at low-energy or long-distance and long-time scales in systems with conserved quantities such as charge and spin (see Refs.~\cite{landau,Kovtun:2012rj,Romatschke:2017ejr}). The simplest type of dissipative (or entropy-generating) transport is diffusion, which is described by the following differential equation:
\begin{equation}
    \partial_t \varepsilon (t,{\bf x}) = D \nabla^2 \varepsilon(t,{\bf x}),
\end{equation}
where $\varepsilon$ is the energy density and $D$ the diffusivity (often called the diffusion constant). The momentum space dispersion relation is then given by 
\begin{equation}\label{Diff_DispRel}
    \omega = - i D k^2,
\end{equation}
where $\omega$ is the frequency, $\textbf{k}$ is the wavevector of the mode and, where not stated otherwise, $k= \vert\mathbf{k}\vert$. The retarded two-point function $G_R(t,{\bf x}; t', {\bf x}' ) = \theta(t-t')\langle [\varepsilon(t,{\bf x}), \varepsilon(t',{\bf x}') ]\rangle  $  determines the `linearised' real-time response in such a system. In momentum space, this correlation function is 
\begin{equation}\label{GR_class}
    G_R(\omega, k) = \frac{\chi D k^2}{i\omega - D k^2},
\end{equation}
where $\chi$ is the susceptibility given by $\chi = \varepsilon + p$, with $p$ the pressure. The denominator has a simple pole $\omega - \omega(k)$ with the dispersion relation $\omega(k)$ given by Eq.~\eqref{Diff_DispRel}. In position space, such a simple pole leads to exponential damping in time for real $k$. 

More generally, the diffusive correlator can be derived from a theory of classical first-order hydrodynamics, where `first-order' refers to the order to which the conserved quantity (e.g., a conserved current or the energy-momentum tensor) is expanded in terms of gradients. In principle, this expansion can continue to infinite order in derivatives, leading to the diffusive dispersion relation $\omega(k)$ given as an infinite power series in $k^2$ (see e.g.~Refs.~\cite{Grozdanov:2019kge,Grozdanov:2019uhi} and \cite{Bu:2014ena,Heller:2020uuy}, and references therein). Using the concept of the gradient expansion, but at the level of the action, a comprehensive theory of hydrodynamics formulated as a closed-time-path (CTP) (or Schwinger-Keldysh) effective field theory (EFT) has been developed in recent years \cite{Grozdanov:2013dba,Crossley:2015evo,Haehl:2015uoc,Jensen:2017kzi,Liu:2018kfw,Bu:2020jfo}.
In this paper, we aim to make generic statements about any system that, classically, exhibits diffusion to first order in derivatives, and discuss the effects that the loops of hydrodynamic fields can have on low-energy `diffusive' transport.

The inclusion of loop effects (which traditionally arise from thermal or stochastic fluctuations) leads to the well-known phenomenon of long time tails (see Refs.~\cite{Kovtun:2003vj,Kovtun:2012rj} and also \cite{Caron-Huot:2009kyg,Abbasi:2021fcz}) whereby $G_R(\omega,k)$ ceases to be a meromorphic function and develops non-analyticities with branch cuts. Those give rise to power-law effects, which persist for longer time. From the point of view of the EFT, these effects arise due to the loop effects correcting the classical (tree-level) correlator \eqref{GR_class} to have the form
\begin{equation}\label{GR_loop}
    G_R(\omega, k) = \frac{\left(\chi+\delta \chi(\omega, k)\right) D k^2}{i\omega - D k^2 + i \Sigma(\omega, k)},
\end{equation}
where $\delta\chi$ is a shifted, renormalised susceptibility and $\Sigma$ is the self-energy. The one-loop structure of the theory was investigated in Ref.~\cite{Chen-Lin:2018kfl}, confirming previously known results that the correlator develops non-analyticities (see also \cite{Jain:2020hcu}). In particular, $G_R(\omega,k)$ now has a branch point at $\omega = - i D k^2 / 2$.

\begin{figure}
\centering
\includegraphics[width=0.9\linewidth]{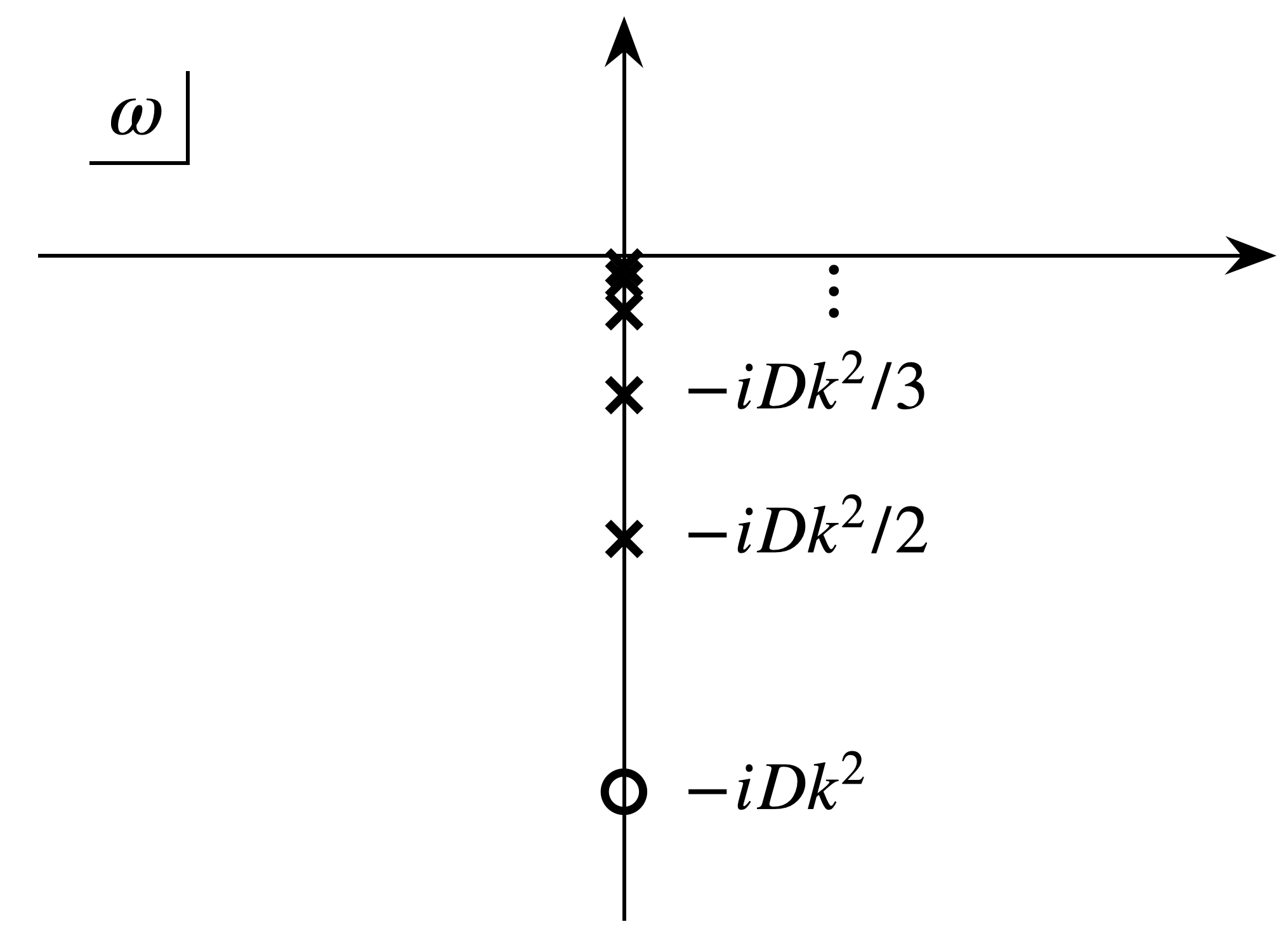}
\caption{Analytic structure of the all-order diffusive retarded correlator \eqref{GR_loop} in the complex $\omega$-plane proposed in Ref.~\cite{Delacretaz:2020nit}. Crosses denote the infinite sequence of branch points \eqref{BranchPoint}, while the empty circle indicates the diffusive pole of the tree-level correlator \eqref{Diff_DispRel}.}
\label{fig:omega plane branch points}
\end{figure}

It is then natural to ask what the generic {\em all}-loop structure of $G_R(\omega,k)$ is. In Ref.~\cite{Delacretaz:2020nit}, Delacr\'{e}taz discussed this problem by using generic conformal field theory (CFT) arguments to understand the decay of hydrodynamic modes into UV excitations at late times. Without reliance on a hydrodynamic EFT, he argued that an $n$-loop contribution will generically introduce new branch cuts with a branch point at
\begin{equation}\label{BranchPoint}
    \omega = - \frac{iD k^2}{n+1}.  
\end{equation}
This therefore suggests that the full `perturbative' retarded diffusive correlator $G_R(\omega,k)$ contains an infinite sequence of branch points:
\begin{equation}
    - \frac{iD k^2}{2} , \, - \frac{i D k^2}{3}, \, - \frac{i D k^2}{4} , \, \ldots,
    \label{branch point sequence}
\end{equation}
limiting towards the origin $\omega = 0$ along the imaginary axis (for $k \in \mathbb{R}$). We schematically depict this analytic structure in the complex $\omega$-plane in Figure~\ref{fig:omega plane branch points}. Complementary to this discussion, a more recent work \cite{Michailidis:2023mkd} computed the hydrodynamic two-loop structure in charge-conjugation-symmetric states (with a vanishing one-loop contribution), confirming the proposal in Eq.~\eqref{BranchPoint} at two loops. In this paper, we re-analyse and expand  on the previously known one- and two-loop results and then, by evaluating a subset of Feynman diagrams for all loops, known as {\it banana diagrams}, provide a general argument that confirms this expectation directly within the framework of hydrodynamic EFT. Analysing this particular subset of all diagrams can be motivated by `reasonable' physical assumptions that we make about the scalings of the all-order couplings. Finally, we also investigate a number of simple features that the all-loop result implies.

This paper is structured as follows: in Section~\ref{sec:EFT}, we write down the complete EFT of diffusion to first order in gradient expansion. In Sections~\ref{sec:oneloop} and~\ref{sec:twoloop}, we summarize the one-loop and two-loop corrections to the classical results, respectively, and in Section~\ref{sec:nloopbranchcut}, we present the argument for Eq.~\eqref{BranchPoint}. We conclude with the analysis of the $n$-loop result within the sector of banana diagrams in Section~\ref{sec:nloopbranchcutAnalysis}. Finally, two Appendices are also added that contain certain details of the one-loop calculation.

\section{Complete EFT of diffusion}\label{sec:EFT}

We begin by briefly outlining the ingredients of the effective closed-time-path (CTP) field theory of diffusion. For details, see Refs.~\cite{Liu:2018kfw,Chen-Lin:2018kfl}. The path integral defining the theory in the state with density matrix $\rho$ has the form 
\begin{equation}
Z(A_{i1},A_{j2})=\int_\rho D\alpha_{i1} D\alpha_{j2}\,e^{iS(\alpha_{i1},A_{i1})-iS(\alpha_{j2},A_{j2})},
\label{Zpathint}
\end{equation}
where $\alpha_{i1,j2}$ denotes the doubled field located on two parts of the time contour and $S(\alpha_i,A_i)=S(\alpha_i)+\int\dd x\,\mathcal{O}_i(x)A_i(x)$, where $A_{i1,j2}$ are the sources of some doubled field theory operators $\mathcal{O}_{i1,j2}$. Assuming a separation of scales, we can rewrite Eq.~\eqref{Zpathint} in terms of a different set of doubled (hydrodynamic) degrees of freedom $\varphi_{i1,j2}$:
\begin{equation}
Z(A_{i1},A_{j2}) = \int D\varphi_{i1} D\varphi_{j2}\,e^{iS_\text{eff}(\varphi_{i1},A_{i1},\varphi_{j2},A_{j2})},
\end{equation}
where $S_\text{eff}(\varphi_{i1},A_{i1},\varphi_{j2},A_{j2})$ is the effective action. 

It is now convenient to introduce the so-called Keldysh basis:
\begin{equation}
\varphi_{ir}=\frac{1}{2}(\varphi_{i1}+\varphi_{i2}),\quad\quad\varphi_{ia} = \varphi_{i1}-\varphi_{i2},
\end{equation}
in which the effective CTP action $S_\text{eff}$ must statisfy the following standard set of conditions:
\begin{align}
&S_\text{eff}(\varphi_{ir},A_{ir},\varphi_{ia},A_{ia})^*=\nonumber\\
&\phantom{S_\text{eff}(}-S_\text{eff}(\varphi_{ir},A_{ir},-\varphi_{ia},-A_{ia}),\label{Icondconj}\\
&S_\text{eff}(\varphi_{ir},A_{ir},\varphi_{ia}=0,A_{ia}=0)=0,\label{Iconda0}\\
&\mathrm{Im}(S_\text{eff})\geq 0.
\end{align}

Setting the sources to zero, we can decompose the effective action into free and interacting parts:
\begin{align}
S_\text{eff} &= \int\dd^{d+1}x \left(\mathcal{L}_2 + \mathcal{L}_{\mathrm{int}}\right),\\
\mathcal{L}_2 &= \frac{i}{2}H_{ij}\varphi_{ia}\varphi_{ja}+K_{ij}\varphi_{ia}\varphi_{jr}.
\end{align}
Here, $\mathcal{L}_2$ denotes the free part with $H_{ij}$ and $K_{ij}$ the lowest-derivative-order differential operators. The term $\varphi_{ir}\varphi_{jr}$ is forbidden due to the condition in Eq.~\eqref{Iconda0}. We find the propagators by computing the inverse matrix
\begin{equation}
\begin{pmatrix}
H & -iK\\
-iK^\dagger & 0
\end{pmatrix}^{\!-1}=
\begin{pmatrix}
0 & -iK^{-\dagger}\\
iK^{-1} & K^{-1}HK^{-\dagger}
\end{pmatrix}, \label{2pointmatrix}
\end{equation}
where $K^{-\dagger}=(K^\dagger)^{-1}$.

The `macroscopic' operators $\mathcal{O}$ we are interested in can be formally expressed in terms of $\varphi$ as
\begin{equation}
\mathcal{O} = \sum_{n=1}^\infty \sum_{k = 1}^{m_n}\prod_{l=1}^n D_{nkli}\varphi_i,
\end{equation}
where $D_{nkli}$ is a differential operator and $n$ is the order in the fields $\varphi_i$. In particular, we can take $m_1=1$ to see that
\begin{align}
\mathcal{O}_r &= D_{111i}\varphi_{ir}+\ldots,\nonumber\\
\mathcal{O}_a &= D_{111i}\varphi_{ia}+\ldots, \label{Oraexp}
\end{align}
and, more generally, we have
\begin{align}
&\prod_{l=1}^n D_{nkli}\varphi_{i1} - \prod_{l=1}^n D_{nkli}\varphi_{i2} 
\nonumber\\&=\prod_{l=1}^n D_{nkli}\bigg(\varphi_{ir} + \frac{1}{2}\varphi_{ia}\bigg) - \prod_{l=1}^n D_{nkli}\bigg(\varphi_{ir} - \frac{1}{2}\varphi_{ia}\bigg).
\end{align}
We can then perturbatively write the retarded correlator of $\mathcal{O}$ as the following expansion: 
\begin{align}
&\langle\mathcal{P}(\mathcal{O}_r(x)\mathcal{O}_a(y))\rangle_\rho = \langle D_{111i}\varphi_{ir}(x)D_{111j}\varphi_{ja}(y)\rangle+\ldots\nonumber\\
&=iD_{111i}\vert_x D_{111j}\vert_y (K^{-1})_{ij}(x-y)+\ldots .
\end{align}
Here, the averaging should be understood as done with the path integral \eqref{Zpathint}. Thus, we can expand any retarded correlator $\mathcal{O}_{r,a}$ into a combination of $\varphi_{ir}$ and $\varphi_{ia}$, and compute it perturbatively using the propagators \eqref{2pointmatrix}. We note that when computing the Feynman diagrams associated with the hydrodynamic correlators, we will typically encounter divergences in the momentum integrals. For this reason, we can introduce a hard momentum cut-off $k < k_{\mathrm{max}}$ and regularise the frequency by taking
\begin{equation}
K_{ij}(\omega,k)\longrightarrow K_{ij}(\omega,k)\bigg(\frac{\Lambda-i\omega}{\Lambda}\bigg)^{\! r},\label{regularization}
\end{equation}
where $\Lambda$ is the regulator and $r$ is an exponent, which we choose so that the integral is convergent.

We start with an effective action $S_\text{eff}(A_{a\mu} + \partial_\mu \varphi_a,A_{r\mu} + \partial_\mu \varphi_r)$, which we study perturbatively to all orders in the powers of the hydrodynamic fields, and at leading order in the number of derivatives acting of the fields. We write
\begin{equation}\label{Seff_exp_n}
S_\text{eff} = \int \dd^{d+1}x\, \left(\mathcal{L}_2 + \mathcal{L}_3 + \ldots\right),
\end{equation}
where $\mathcal{L}_n$ contains $n$ powers of the fields. In what follows, we will set $A_\mu=0$. We will also take the action to be invariant under the symmetry $\varphi_r(t,{\bf x})\rightarrow\varphi_r(t,{\bf x}) + \lambda({\bf x})$. Moreover, we note that since we are interested in a theory of diffusion, we can choose to count the `sizes' of various derivatives that act on the hydrodynamic fields as $\partial_t\sim\nabla^2$. 
 
As a result of the condition in Eq.~\eqref{Iconda0}, each possible term will contain a factor of $\varphi_a$. The possible quadratic terms are then,
\begin{equation}
\dot{\varphi}_a^2,\quad\quad \dot{\varphi}_r\dot{\varphi}_a,\quad\quad \dot{\varphi}_r\nabla^2\varphi_a,\quad\quad (\nabla\varphi_a)^2.
\end{equation}
We identify $\varepsilon$ with the functional derivative 
\begin{equation}
\varepsilon = \frac{\delta S_\text{eff}}{\delta A_{a0}}\bigg\vert_{A_a=A_r=0}\label{epsilonfuncder},
\end{equation}
which allows us to discard the $\dot{\varphi}_a^2$ term since, to linear order, $\varepsilon\sim \dot\varphi_r$. One can now replace all $\dot\varphi_r$ with $\varepsilon$ and ignore the time derivatives on the $\varphi_r$ fields. Thus, the most general Lagrangian can be parameterised as
\begin{equation}
\sum_{n=2}^\infty\mathcal{L}_n=\varepsilon \left(\dot{\varphi}_a + f_1(\varepsilon)\nabla^2\varphi_a\right) + if_2(\varepsilon)(\nabla\varphi_a)^2, \label{generalL}
\end{equation}
where $f_{1,2}$ are real and analytic functions of $\varepsilon$. In light of the effective action expansion \eqref{Seff_exp_n}, we will therefore consider $f_{1,2}$ as series expanded in powers of $\varepsilon$.

Let us first consider the quadratic part of the effective CTP action, which is given by
\begin{equation}
\mathcal{L}_2 = iT^2\kappa(\nabla\varphi_a)^2 - \varphi_a\left(\dot{\varepsilon}- D\nabla^2\varepsilon\right), \label{L2}
\end{equation}
where $T$ is the temperature, $\kappa$ is the thermal conductivity and $D$ is the diffusion constant. 
Working in Fourier space
with $p=(\omega,\mathbf{k})$, we can rewrite the quadratic action as
\begin{align}\nonumber
iS_\text{eff}^{(2)} &= -\frac{1}{2}\int \dd^{d+1}p\,
\begin{pmatrix}
\varphi_a(p)^* & \varepsilon(p)^*
\end{pmatrix}\\
&\times
\begin{pmatrix}
2T^2\kappa k^2 & \omega + iDk^2\\
-\omega+iDk^2 & 0
\end{pmatrix}
\begin{pmatrix}
\varphi_a(p)\\
\varepsilon(p)
\end{pmatrix}.
\end{align}
This expression determines all of the CTP propagators. Denoting $\varepsilon$ with a bold line and $\varphi$ with a wavy line, the propagators are given by
\begin{align}
\gef(p)\quad &=\quad 
\begin{tikzpicture}
    \begin{scope}[very thick]
    \draw[-{Stealth[width=4mm, length=2mm]}, shorten >=-1.6pt] (0,0)--(1,0);
    \draw[] (1,0)--(1,0);
    \draw[snek] (1,0)--(2,0);
    \node at (1.1,.3) {$p$};
    \end{scope}
    \end{tikzpicture}\quad =\quad \frac{1}{\omega + iDk^2},\nonumber\\
\gfe(p)\quad &=\quad 
\begin{tikzpicture}
    \begin{scope}[very thick]
    \draw[snek, -{Stealth[width=4mm, length=2mm]}, shorten >=-1.6pt] (0,0)--(1,0);
    \draw[] (1,0)--(2,0);
    \node at (1.1,.3) {$p$};
    \end{scope}
    \end{tikzpicture}\quad =\quad \frac{1}{-\omega + iDk^2},\nonumber\\
\gee(p)\quad &=\quad 
\begin{tikzpicture}
    \begin{scope}[very thick]
    \draw[-{Stealth[width=4mm, length=2mm]}, shorten >=-1.6pt] (0,0)--(1,0);
    \draw[] (1,0)--(2,0);
    \node at (1.1,.3) {$p$};
    \end{scope}
    \end{tikzpicture}\quad =\quad \frac{2T^2\kappa k^2}{\omega^2 + D^2 k^4},\nonumber\\
G_{\varphi\varphi}(p)\quad &=\quad 
\begin{tikzpicture}
    \begin{scope}[very thick]
    \draw[snek, -{Stealth[width=4mm, length=2mm]}, shorten >=-1.6pt] (0,0)--(1,0);
    \draw[snek] (1,0)--(2,0);
    \node at (1.1,.3) {$p$};
    \end{scope}
    \end{tikzpicture}\quad =\quad 0.\label{props}
\end{align}

\begin{figure*} 
\begin{alignat}{4} 
\begin{tikzpicture}[baseline]
    \begin{scope}[very thick]
    \draw[snek, -{Stealth[width=4mm, length=2mm]}, shorten >=-1.6pt] (0,0)--(1,0);
    \draw[-{Stealth[width=4mm, length=2mm]}, shorten >=-1.6pt] (0,0)--(-.5,.5 * 1.732);
    \draw[-{Stealth[width=4mm, length=2mm]}, shorten >=-1.6pt] (0,0)--(-.5,-.5 * 1.732);
    \node at (.9, .4) {$p$};
    \end{scope}
\end{tikzpicture}
&\begin{tikzpicture}
    \node at (0,.05) {$=i\lambda_0 \mathbf{k}^2$};
\end{tikzpicture}
%%%%%%%%%%%%%%%%%%%%%%%%%%%%%%%%%% first
\quad
\begin{tikzpicture}[baseline]
    \begin{scope}[very thick]
    \draw[snek, -{Stealth[width=4mm, length=2mm]}, shorten >=-1.6pt] (0,0)--(1,0);
    \draw[snek, -{Stealth[width=4mm, length=2mm]}, shorten >=-1.6pt] (0,0)--(-.5,.5 *1.732);
    \draw[-{Stealth[width=4mm, length=2mm]}, shorten >=-1.6pt] (0,0)--(-.5,-.5 * 1.732);
    \node at (-.75, .8) {$p'$};
    \node at (.9, .4) {$p$};
    \end{scope}
\end{tikzpicture}
&&\begin{tikzpicture}	
	\node at (0,.05) {$=2cT^2\tilde\lambda_0 \mathbf{k} \cdot \mathbf{k}'$};
\end{tikzpicture}\quad 
%%%%%%%%%%%%%%%%%%%%%%%%%%%%%%%%%%%% second
\begin{tikzpicture}[baseline]
    \begin{scope}[very thick]
    \draw[snek, -{Stealth[width=4mm, length=2mm]}, shorten >=-1.6pt] (0,0)--(1,0);
    \draw[-{Stealth[width=4mm, length=2mm]}, shorten >=-1.6pt] (0,0)--(0,1);
    \draw[-{Stealth[width=4mm, length=2mm]}, shorten >=-1.6pt] (0,0)--(0,-1);
    \draw[-{Stealth[width=4mm, length=2mm]}, shorten >=-1.6pt] (0,0)--(-1,0);
    \node at (.9, .4) {$p$};
    \end{scope}
\end{tikzpicture}
&&&\begin{tikzpicture}	
	\node at (0,.05) {$=2i\lambda_1\mathbf{k}^2$};
\end{tikzpicture}\quad
%%%%%%%%%%%%%%%%%%%%%%%%%%%% third
\begin{tikzpicture}[baseline]
    \begin{scope}[very thick]
    \draw[snek, -{Stealth[width=4mm, length=2mm]}, shorten >=-1.6pt] (0,0)--(1,0);
    \draw[snek, -{Stealth[width=4mm, length=2mm]}, shorten >=-1.6pt] (0,0)--(0,1);
    \draw[-{Stealth[width=4mm, length=2mm]}, shorten >=-1.6pt] (0,0)--(0,-1);
    \draw[-{Stealth[width=4mm, length=2mm]}, shorten >=-1.6pt] (0,0)--(-1,0);
    \node at (.9, .4) {$p$};
    \node at (-.35, .8) {$p'$};
    \end{scope}
\end{tikzpicture}
&\begin{tikzpicture}	
	\node at (0,.05) {$=4cT^2\tilde\lambda_1 \mathbf{k} \cdot \mathbf{k}'$};
\end{tikzpicture}
%%%%%%%%%%%%%%%%%%%%%%%%%%%%%%%%%% fourth, line break
\nonumber \\
\begin{tikzpicture}[baseline]
    \begin{scope}[very thick]
    \draw[snek, -{Stealth[width=4mm, length=2mm]}, shorten >=-1.6pt] (0,0)--(1,0);
    \draw[-{Stealth[width=4mm, length=2mm]}, shorten >=-1.6pt] (0,0)--({cos(72)}, {sin(72)} );
    \draw[-{Stealth[width=4mm, length=2mm]}, shorten >=-1.6pt] (0,0)--({cos(72)}, -{sin(72)} );
    \draw[-{Stealth[width=4mm, length=2mm]}, shorten >=-1.6pt] (0,0)--({cos(144)}, {sin(144)} );
    \draw[-{Stealth[width=4mm, length=2mm]}, shorten >=-1.6pt] (0,0)--({cos(144)}, -{sin(144)} );
    \node at (.9, .4) {$p$};
    \end{scope}
\end{tikzpicture}
&\begin{tikzpicture}	
	\node at (0,.05) {$=6i\lambda_2\mathbf{k}^2$};
\end{tikzpicture}
%%%%%%%%%%%%%%%%%%%%%%%%%%%%%% fifth
\quad
\begin{tikzpicture}[baseline]
    \begin{scope}[very thick]
    \draw[snek, -{Stealth[width=4mm, length=2mm]}, shorten >=-1.6pt] (0,0)--(1,0);
    \draw[snek, -{Stealth[width=4mm, length=2mm]}, shorten >=-1.6pt] (0,0)--({cos(72)}, {sin(72)} );
    \draw[-{Stealth[width=4mm, length=2mm]}, shorten >=-1.6pt] (0,0)--({cos(72)}, -{sin(72)} );
    \draw[-{Stealth[width=4mm, length=2mm]}, shorten >=-1.6pt] (0,0)--({cos(144)}, {sin(144)} );
    \draw[-{Stealth[width=4mm, length=2mm]}, shorten >=-1.6pt] (0,0)--({cos(144)}, -{sin(144)} );
    \node at (.9, .4) {$p$};
    \node at (-.1, .8) {$p'$};
    \end{scope}
\end{tikzpicture}
&&\begin{tikzpicture}	
	\node at (0,.05) {$=12cT^2\tilde\lambda_2 \mathbf{k}\cdot \mathbf{k}'$};
\end{tikzpicture}
%%%%%%%%%%%%%%%%%%%%%%%%%%%%%%% sixth
\quad 
\begin{tikzpicture}[baseline]
    \begin{scope}[very thick]
    \draw[snek, -{Stealth[width=4mm, length=2mm]}, shorten >=-1.6pt] (0,0)--(1,0);
    \draw[-{Stealth[width=4mm, length=2mm]}, shorten >=-1.6pt] (0,0)--({cos(60)}, {sin(60)} );
    \draw[-{Stealth[width=4mm, length=2mm]}, shorten >=-1.6pt] (0,0)--({cos(120)}, {sin(120)} );
    \draw[-{Stealth[width=4mm, length=2mm]}, shorten >=-1.6pt] (0,0)--({cos(180)}, {sin(180)} );
    \draw[-{Stealth[width=4mm, length=2mm]}, shorten >=-1.6pt] (0,0)--({cos(240)}, {sin(240)} );
    \draw[-{Stealth[width=4mm, length=2mm]}, shorten >=-1.6pt] (0,0)--({cos(300)}, {sin(300)} );
    \node at (.9, .4) {$p$};
    \end{scope}
\end{tikzpicture}
&&&\begin{tikzpicture}	
	\node at (0,.05) {$=24i\lambda_3\mathbf{k}^2$};
\end{tikzpicture}\quad
%%%%%%%%%%%%%%%%%%%%%%%%%%%% sixth
\begin{tikzpicture}[baseline]
    \begin{scope}[very thick]
    \draw[snek, -{Stealth[width=4mm, length=2mm]}, shorten >=-1.6pt] (0,0)--(1,0);
    \draw[snek, -{Stealth[width=4mm, length=2mm]}, shorten >=-1.6pt] (0,0)--({cos(60)}, {sin(60)} );
    \draw[-{Stealth[width=4mm, length=2mm]}, shorten >=-1.6pt] (0,0)--({cos(120)}, {sin(120)} );
    \draw[-{Stealth[width=4mm, length=2mm]}, shorten >=-1.6pt] (0,0)--({cos(180)}, {sin(180)} );
    \draw[-{Stealth[width=4mm, length=2mm]}, shorten >=-1.6pt] (0,0)--({cos(240)}, {sin(240)} );
    \draw[-{Stealth[width=4mm, length=2mm]}, shorten >=-1.6pt] (0,0)--({cos(300)}, {sin(300)} );
    \node at (.9, .4) {$p$};
    \node at (.1, .8) {$p'$};
    \end{scope}
\end{tikzpicture}
&\begin{tikzpicture}	
	\node at (0,.05) {$=48cT^2\tilde\lambda_3\mathbf{k}\cdot \mathbf{k}'$};
\end{tikzpicture}\nonumber
\end{alignat}
\caption{The vertices of the theory that follow from the Lagrangian \eqref{generalL}. We use $p=(\omega,\mathbf{k})$ to denote the spacetime momenta.}\label{fig:vertices}
\end{figure*}
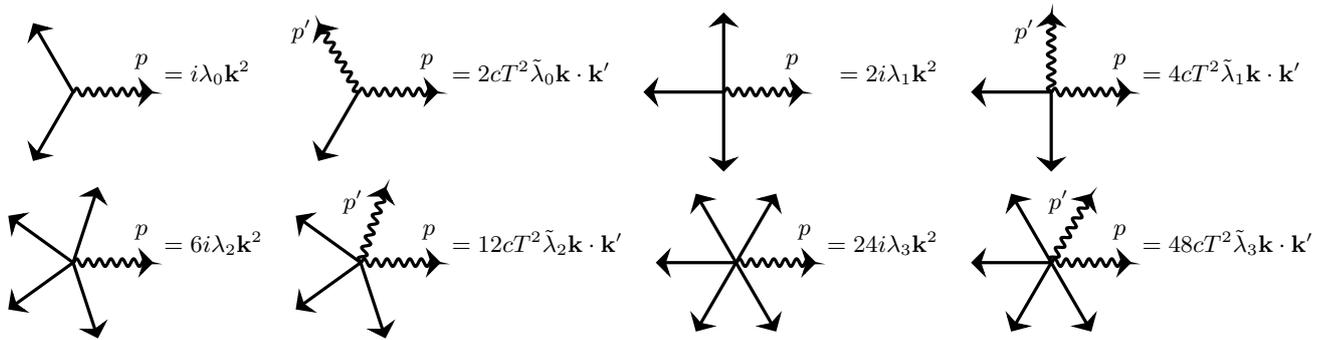

The vertices of the theory are extracted by including higher-order terms from the Lagrangian \eqref{generalL}. Namely, we expand it as 
\begin{align}
\mathcal{L}_{\text{int}}&=\nabla^2\varphi_a \left(\frac{1}{2}\lambda_0\varepsilon^2 + \frac{1}{3}\lambda_1\varepsilon^3 + \frac{1}{4}\lambda_2\varepsilon^4 + \frac{1}{5}\lambda_3\varepsilon^5 \right) \nonumber\\
&+ icT^2(\nabla\varphi_a)^2\left(\tilde\lambda_0\varepsilon+\tilde\lambda_1\varepsilon^2 + \tilde\lambda_2\varepsilon^3 + \tilde\lambda_3\varepsilon^4\right)+\ldots,
\label{Lint}
\end{align}
where $c=\kappa/D$ is the specific heat. Different vertices are depicted in Figure~\ref{fig:vertices}.  

We are interested in the higher-loop corrections to the retarded propagator $G_R$, which are easiest to compute from the symmetrised propagator $G_S$. The relationship between $G_R$ and $G_S$ is given by the following formula:
\begin{align}
    \text{Im} \, G_R = \tanh\left(\frac{\omega}{2T}\right) G_S,
    \label{GR_GS_relation}
\end{align}
from which we read off the contribution to the self-energy. We can calculate $G_S$ by introducing the one-particle-irreducible diagrams $C(\omega, k)$ and $\Sigma(\omega, k)$. They enter into the expression for $G_S$ through the Schwinger-Dyson series as 
\begin{align}
    G_S &= -G_{\varepsilon\varphi} C G_{\varphi\varepsilon} \sum_{m,n=0}^\infty \left( -G_{\varepsilon\varphi} \Sigma \right)^m \left( G_{\varphi\varepsilon} \Sigma^* \right)^n \nonumber \\
    &= \frac{C}{\omega^2 + D^2 k^4 + 2\omega \text{Re}(\Sigma) + 2Dk^2\text{Im}(\Sigma) + |\Sigma|^2},
    \label{GsymExpression}
\end{align}
where $C$ denotes the diagrams that attach to an external field $\varphi$ on both ends, while the $\Sigma$ diagrams connect to a $\varphi$-line on one side and an $\varepsilon$-line on the other.

\section{One-loop structure}\label{sec:oneloop}

We can now compute the one-loop corrections to $\delta\chi(\omega, k)$ and $\Sigma(\omega, k)$ that enter the retarded correlator $G_R(\omega, k)$ by computing all of the one-loop contributions to the functions $C$ and $\Sigma$ (cf.~Eq.~\eqref{GsymExpression}). At this order, the only two structurally different one-loop diagrams that enter into the calculation have the form
\begin{equation}
\begin{tikzpicture}
\begin{scope}[very thick]
\draw[] ((-.75 - 2 * .75 *pi/4,0)--(-.75,0);
\draw[] (.75 + 2 * .75*pi/4,0)--(.75,0);
\centerarc[](0,0)(0:360:.75)
\end{scope}
\end{tikzpicture}
\quad
\begin{tikzpicture}
\begin{scope}[very thick]
\draw[] (-.75 - 2 * .75 *pi/4,-.75)--(.75 + 2 * .75*pi/4,-.75);
\centerarc[](0,0)(0:360:.75)
\end{scope}
\end{tikzpicture}
\nonumber
\end{equation}
where each solid line in the above figure stands for a generic propagator, i.e., any one of the propagators stated in Eq.~\eqref{props}. The calculation was done in Ref.~\cite{Chen-Lin:2018kfl}, so we only state the results here and, for the sake of completeness, show further details of the calculation in Appendix \ref{app:details1loop}. We also note that, in this work, beyond the one-loop analysis, we will only focus on the analytic structure of the denominator of $G_R$, which therefore only requires a detailed analysis of $\Sigma$ and not $C$. 

To first order in momenta and to first order in non-analyticities, the two corrections to the retarded propagator in \eqref{GR_loop} turn out to be of the form
\begin{align}
    \Sigma(\omega, k) &= i \delta D k^2 + \Sigma_{\star}(\omega, k), 
    \label{sigma1loop}\\
    \delta \chi(\omega, k) &= \delta\chi_0 + \chi_{\star}(\omega, k),
    \label{deltachi1loop}
\end{align}
where $\star$ denotes the non-analytic terms, while $\delta D$ and $\delta\chi_0$ contain all analytic terms that depend on the cut-off and are momentum-independent. Physically, $\delta D$ represents a correction to the diffusion constant and $\delta \chi_0$ to the susceptibility. It is important to keep in mind that there is no way to remove the cut-off dependence from the calculation. 

Referring again to Appendix \ref{app:details1loop} for the full calculation and exact expressions, the final result indicates that, instead of a single diffusive pole, the retarded correlator now exhibits a pair of poles at
\begin{align}
    \omega = -i D' k^2 \pm \delta\omega(k),
    \label{smallwkpoleassumption}
\end{align}
with a renormalised diffusivity $D' = D+\delta D$ and a `small' sub-leading\footnote{Note that the Lagrangian \eqref{Lint} is expanded in terms of $\varepsilon$, where the coefficients of the expansion are denoted by $\lambda_i$ and $\tilde\lambda_i$. It is in the context of this small $\varepsilon$ expansion, that the two corrections, \eqref{sigma1loop} and \eqref{deltachi1loop}, are subleading.} shift:
\begin{align}
    \delta \omega(k) \propto \gamma^{d/2-1}k^{d+2}\begin{cases}
        1 & d\text{ odd}, \\
        \log(\gamma k^2) & d\text{ even}.
    \end{cases}
    \label{delta omega 1 loop}
\end{align}
In the above expression, we also include a factor of 
\begin{equation}
    \gamma = 1-\frac{2D'}{D},
    \label{gamma}
\end{equation}
which determines whether the tree-level diffusive pole will split in the real or imaginary $\omega$ direction. For example, in $d=3$ and for negative $\gamma$, the splitting occurs in the real direction (the pair of poles acquires a real part signaling propagation), while if $\gamma>0$, the splitting occurs in the imaginary direction (the pair of poles remains purely relaxing).

Besides the two poles, the retarded correlator also develops a branch point at
\begin{align}
    \omega = -\frac{iD k^2}{2},
    \label{branchPoint}
\end{align}
where the exact structure of the Riemann surface depends on the number of spatial dimensions $d$. This is because the two non-analytic corrections are both of the form
\begin{align}
    \chi_\star, \Sigma_\star \propto k^2 z^{d-2} \begin{cases}
        1 & d\text{ odd}, \\
        \log(z^2) & d\text{ even},
    \end{cases}
\end{align}
where 
\begin{equation}
    z^2 = k^2 - \frac{2i\omega}{D}.
\end{equation}

The EFT of hydrodynamic diffusion \eqref{generalL} is a theory valid only for small $\omega$ and $k$ as compared to the equilibrium temperature and the cut-off of the theory. However, here, we go beyond its strict regime of validity and study the one-loop corrected analytic structure `non-perturbatively' in $\omega$ and $k$. The reasons for this are two-fold. The first is a pure academic interest in the intricate mathematical analytic structures of two-point correlators that `simple' EFTs can exhibit. The second is more physically motivated and is related to the recent advances in understanding classical (tree-level) relativistic hydrodynamics in an arbitrary frame that allows for diffusion to be made stable and causal for large $\omega$ and $k$ as a result of an analogously  non-perturbative treatment of the theory that naturally gives rise to new gapped poles of $G_R$ (see the BDNK theory in Refs.~\cite{Bemfica:2017wps, Kovtun_2019}). While it is unclear whether similar `non-perturbative' treatments of loop corrections can be useful in stabilising hydrodynamic simulations, we nevertheless proceed with the hope that such investigations may prove useful in the future. 

\begin{figure*}[ht!]
\centering
\includegraphics[]{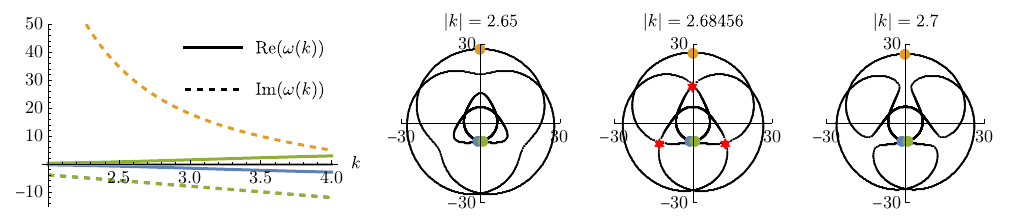}
\caption{Poles of the one-loop retarded correlator in the case of $D'>D/2$. (We set $c$, $T$, $D$, $D'$ and $\lambda_1$, $\tilde{\lambda}_1$ all equal to one.) On the left, we plot the real and imaginary parts of their dispersion relation, with solid and dashed lines, respectively, as functions of real $k$. In the right three plots, we show the trajectories of the poles in the complex $\omega$-plane at fixed $|k|$ with varying $\text{Arg}[k] \in [0,2\pi)$. We choose the values of $|k|$ just before, at, and after the value that results in the pair-wise collisions of the three modes. Dots with three colours signify the values of the dispersion relations at real $k$ and correspond to the same colours chosen on the left-most plot. Red stars denote the locations of the collisions among the poles (the critical points).}
\label{fig:n1below}
\end{figure*}

\begin{figure*}
\centering
\includegraphics[]{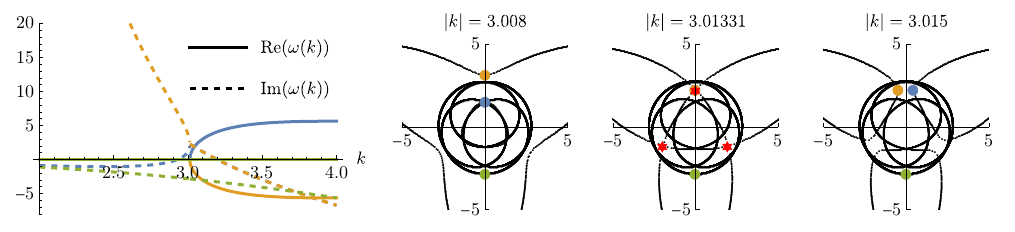}
\caption{Poles of the one-loop retarded correlator in the case of $D'< D/2$. (We set $D'=1/4$, all other parameters are set to one.) In this figure, we plot all the same features as in Figure~\ref{fig:n1below} with the same colour-coding. Note that for $k$ chosen at the collision, two of poles coincide for real $k$.}
\label{fig:n1above}
\end{figure*}

In this manner, the one-loop corrected diffusive $G_R$ can be treated as a `full' result. It is then particularly interesting to analyse the behaviour of its poles as a function of complexified $k$, expanding on the considerations of Refs.~\cite{Grozdanov:2019kge,Grozdanov:2019uhi} from classical hydrodynamics. Here, we only focus on the case of $d=3$ spatial dimensions. Solving for the dispersion relations, one obtains, beyond the two hydrodynamic poles in \eqref{smallwkpoleassumption}, a third mode of which the dispersion relation diverges for small $k$. As in the perturbative case (expanded around $\omega = 0$ and $k=0$), it is instructive to split the analysis into two cases, taking $\gamma$ from Eq.~\eqref{gamma} as either positive or negative. The analysis of the critical points of complex spectral curves \cite{Grozdanov:2019kge,Grozdanov:2019uhi,Grozdanov:2021jfw} also allows us to obtain the locations of collisions of the three modes in the complex plane. We plot the behaviour of the poles of $G_R$ in the vicinity of the collision for a choice of $\gamma<0$ in Figure~\ref{fig:n1below}  and for a choice of $\gamma>0$ in Figure~\ref{fig:n1above}.


\section{Two-loop structure}\label{sec:twoloop}

In order to investigate the analytic structure of diffusive correlators at two loops, we begin by enumerating the possible types of diagrams. At this order, we find $5$ families of different Feynman graphs. Using the handshaking lemma for two loops, we get $10$ different diagrams overall (sorted below by the number and type of vertices that appear in them):
\begin{itemize}
\item $4$ trivalent vertices:
\begin{subequations}
\begin{equation}
\begin{tikzpicture}
\begin{scope}[very thick]
\draw[] (-.75 - 2 * .75 *pi/4,0)--(-.75,0);
\draw[] (0,.75)--(0,-.75);
\draw[] (.75 + 2 * .75*pi/4,0)--(.75,0);
\centerarc[](0,0)(0:360:.75)
\end{scope}
\end{tikzpicture}
\end{equation}
\item $2$ tetravalent vertices:
\begin{equation}
\begin{tikzpicture}
\begin{scope}[very thick]
\draw[] (-.75 - 2 * .75 *pi/4,0)--(.75 + 2 * .75*pi/4,0);
\centerarc[](0,0)(0:360:.75)
\end{scope}
\end{tikzpicture}
\qquad
\begin{tikzpicture}
\begin{scope}[very thick]
\draw[] (-1.5,0)--(1.5,0);
\centerarc[](0,.75)(0:360:.75)
\centerarc[](0,1.125)(0:360:.375)
\end{scope}
\end{tikzpicture}
\end{equation}
\item $1$ tetravalent and $2$ trivalent vertices:
\begin{align}
\begin{tikzpicture}
\begin{scope}[very thick]
\draw[] (-1.5, 0)--(-.75, 0);
\draw[] (-.75, 0)--(0, 0);
\draw[] (1.5, 0)--(2.25, 0);
\draw[] (2.25, 0)--(3, 0);
\centerarc[](0, 0)(0:30:1.5)
\centerarc[](0, 0)(30:60:1.5)
\centerarc[](.75, {sqrt(3) * .75})(240:270:1.5)
\centerarc[](.75, {sqrt(3) * .75})(270:300:1.5)
\centerarc[](1.5, 0)(180:150:1.5)
\centerarc[](1.5, 0)(150:120:1.5)
\centerarc[](2.25, {sqrt(3) * .75})(180:210:1.5)
\centerarc[](2.25, {sqrt(3) * .75})(210:240:1.5)
\end{scope}
\end{tikzpicture}
\quad & \
\begin{tikzpicture}
\begin{scope}[very thick]
\draw[] (-1.5,0)--(1.5,0);
\draw[] (-0.75,0)--(0,.75)--(0.75,0);
\centerarc[](0,1.125)(0:360:.375)
\end{scope}
\end{tikzpicture}\nonumber
\\
\begin{tikzpicture}
\begin{scope}[very thick]
\draw[] (-2.25,0)--(-1.125,0);
\draw[] (2.25,0)--(1.125,0);
\centerarc[](-0.5625,0)(0:360:0.5625)
\centerarc[](0.5625,0)(0:360:0.5625)
\end{scope}
\end{tikzpicture}
\quad & \
\begin{tikzpicture}
\begin{scope}[very thick]
\draw[] (-1.5,0)--(1.5,0);
\centerarc[](0,0)(60:120:1.5)
\centerarc[](-.75,{.75 * sqrt(3)})(300:360:1.5)
\centerarc[](.75,{.75 * sqrt(3)})(180:240:1.5)
\centerarc[](0,{1.5 * sqrt(3)})(240:300:1.5)
\end{scope}
\end{tikzpicture}
\end{align}

\item $1$ trivalent and $1$ pentavalent vertex:

\begin{equation}
\begin{tikzpicture}
\begin{scope}[very thick]
\draw[] (-.75 - 2 * .75 *pi/4,0)--(.75 + 2 * .75*pi/4,0);
\centerarc[](0,0)(180:0:.75)
\centerarc[](0.75,-0.375)(-90:270:0.375)
\end{scope}
\end{tikzpicture}
\qquad
\begin{tikzpicture}
\begin{scope}[very thick]
\draw[] (-1.5,-.75)--(1.5,-.75);
\draw[] (0,.75)--(0,-.75);
\centerarc[](0,0)(0:360:.75)
\end{scope}
\end{tikzpicture}  
\end{equation}

\item $1$ hexavalent vertex:
\begin{equation}
\begin{tikzpicture}
\begin{scope}[very thick]
\draw[] (-1.5,0)--(1.5,0);
\centerarc[](0,.375)(90:-270:.375)
\centerarc[](0,-.375)(-90:270:.375)
\end{scope}
\end{tikzpicture}
\end{equation}
\label{2loopdiagramsintext}
\end{subequations}
\end{itemize}
Only three types of the above diagrams are actually non-factorisable. These are:
\begin{equation*}
\begin{tikzpicture}
\begin{scope}[very thick]
\draw[] (-.75 - 2 * .75 *pi/4,0)--(.75 + 2 * .75*pi/4,0);
\centerarc[](0,0)(0:360:.75)
\end{scope}
\end{tikzpicture}\quad
\begin{tikzpicture}
\begin{scope}[very thick]
\draw[] (-.75 - 2 * .75 *pi/4,0)--(-.75,0);
\draw[] (0,.75)--(0,-.75);
\draw[] (.75 + 2 * .75*pi/4,0)--(.75,0);
\centerarc[](0,0)(0:360:.75)
\end{scope}
\end{tikzpicture}
\end{equation*}
\begin{equation}
\begin{tikzpicture}
\begin{scope}[very thick]
\draw[] (-1.5, 0)--(-.75, 0);
\draw[] (-.75, 0)--(0, 0);
\draw[] (1.5, 0)--(2.25, 0);
\draw[] (2.25, 0)--(3, 0);
\centerarc[](0, 0)(0:30:1.5)
\centerarc[](0, 0)(30:60:1.5)
\centerarc[](.75, {sqrt(3) * .75})(240:270:1.5)
\centerarc[](.75, {sqrt(3) * .75})(270:300:1.5)
\centerarc[](1.5, 0)(180:150:1.5)
\centerarc[](1.5, 0)(150:120:1.5)
\centerarc[](2.25, {sqrt(3) * .75})(180:210:1.5)
\centerarc[](2.25, {sqrt(3) * .75})(210:240:1.5)
\end{scope}
\end{tikzpicture}
\label{nonFactorizableTwoLoop}
\end{equation}
All other families of diagrams can be reduced to (products of) one-loop integrals or integrals that give results of the form $A k^2$ for some (regulator-dependent) constant $A$.

In the following two sections, we will focus on the top left two-loop diagram in \eqref{nonFactorizableTwoLoop} and then analogous $n$-loop diagrams. The reason is that these are the types of Feynman diagrams (with an intermediate decay into the maximal number of virtual particles through a single vertex) that produce new branch points \eqref{BranchPoint}. Moreover, their non-analyticities can be evaluated closed-form for any number of loops. The diagrams of this type are known colloquially as {\em banana diagrams} (see e.g.~Ref.~\cite{Mishnyakov:2023wpd}). In fact, restricting our attention to such diagrams can be justified physically by assuming that all couplings $\lambda_i$ and $\tilde\lambda_i$ in \eqref{Lint} are small and of the same order, which makes the discussed diagram dominate over the remaining two in \eqref{nonFactorizableTwoLoop}. This is because the top left diagram is the only diagram with two (not three) vertices. We will return to this limit of the EFT in the next section. The second reason to restrict our attention to the banana diagram at two loops is in relation to examples of diffusion in theories and states with charge-conjugation symmetry discussed in Ref.~\cite{Michailidis:2023mkd}. For either of these two classes of theories, we can now easily write down the $2$-loop self-energy $\Sigma_\star$. The resulting corrections proportional to $\lambda_1$ and $\tilde\lambda_1$ are
\begin{align}
    \Sigma_\star(&\omega, k) = (-1)^{d+1}\frac{ic^2T^4k^2}{(12\pi\sqrt{3})^d d! D} \nonumber \\
    &\times\left[\lambda_1^2\left(\frac{k^2}{6} - \frac{5i\omega}{2D}\right) + \lambda_1\tilde\lambda_1\left(\frac{7k^2}{6} - \frac{3i\omega}{2D}\right)\right] \nonumber\\
    &\times\left(k^2-\frac{3i\omega}{D}\right)^d \log\left(k^2-\frac{3i\omega}{D}\right).
\end{align}

In considering next the three-loop corrections, one finds that the charge-conjugation symmetry of \cite{Michailidis:2023mkd} is not sufficient for restricting the possible Feynman diagrams to only those of banana-type, which (as we will see) are most clearly responsible for the branch points \eqref{BranchPoint}. A counterexample would be a charge-conjugation symmetric three-loop diagram:
\begin{equation}
\begin{tikzpicture}
\begin{scope}[very thick]
\draw[] (-1.5, 0)--(0, 0);
\draw[] (1.5, 0)--(3, 0);
\centerarc[](0, 0)(0:60:1.5)
\centerarc[](.75, {sqrt(3) * .75})(240:300:1.5)
\centerarc[](1.5, 0)(120:180:1.5)
\centerarc[](2.25, {sqrt(3) * .75})(180:240:1.5)
\centerarc[](-0.75, {sqrt(3) * .75})(300:360:1.5)
\end{scope}
\end{tikzpicture}
\label{ThreeLoopCounterExample}
\end{equation}

\section{The $n$-loop branch cut and `banana diffusion'}\label{sec:nloopbranchcut}

Instead of dwelling on the vast complications associated with higher-and-higher-order loop diagrams, here, we restrict our attention to a `minimal' theory of diffusion that nevertheless exhibits the central feature of an all-loop diffusive correlator: the infinite set of branch points \eqref{BranchPoint} first identified in Ref.~\cite{Delacretaz:2020nit}. In particular, we show how this result arises from purely hydrodynamic EFT calculations.

As already noted in Section~\ref{sec:twoloop}, we will study the expanded CTP effective action \eqref{Lint} within a special regime of the couplings. Namely, we assume they are all of the same order (when made appropriately dimensionless) and small. More precisely, from the Lagrangian \eqref{Lint}, we  see that the mass dimension of $[\lambda_n/\lambda_{n-1}]=[\varepsilon]^{-1}$ and, similarly, for $\tilde{\lambda}_n$. Therefore we assume that the scale of all $\lambda_n \langle \varepsilon \rangle \sim \lambda_{n-1}$ and, similarly, for $\tilde \lambda_n$, with $\langle \varepsilon \rangle$ the energy density of the equilibrium (thermal) state. Thus, at each loop, this scaling only retains the diagrams with the smallest number of vertices, which reduces, to leading-order in the powers of couplings, all possible $n$-loop diagrams to the banana diagrams. Physically, these banana diagrams create and annihilate the $(n+1)$-particle intermediate states. For $n=1,2,3,4$, they are of the following type:
\begin{equation}
\begin{tikzpicture}
\begin{scope}[very thick]
\draw[] (-.75 - 2 * .75 *pi/4,0)--(-.75,0);
\draw[] (.75 + 2 * .75*pi/4,0)--(.75,0);
\centerarc[](0,0)(0:360:.75)
\end{scope}
\end{tikzpicture}
\quad
\begin{tikzpicture}
\begin{scope}[very thick]
\draw[] (-.75 - 2 * .75 *pi/4,0)--(.75 + 2 * .75*pi/4,0);
\centerarc[](0,0)(0:360:.75)
\end{scope}
\end{tikzpicture}
\nonumber
\end{equation}
\begin{equation}
\begin{tikzpicture}
\begin{scope}[very thick]
\draw[] (-.75 - 2 * .75 *pi/4,0)--(-.75,0);
\draw[] (.75 + 2 * .75*pi/4,0)--(.75,0);
\centerarc[](0,{-1.5 * sin(60)})(60:120:1.5)
\centerarc[](0,{1.5 * sin(60)})(240:300:1.5)
\centerarc[](0,0)(0:360:.75)
\end{scope}
\end{tikzpicture}
\quad
\begin{tikzpicture}
\begin{scope}[very thick]
\draw[] (-.75 - 2 * .75 *pi/4,0)--(.75 + 2 * .75*pi/4,0);
\centerarc[](0,{-1.5 * sin(60)})(60:120:1.5)
\centerarc[](0,{1.5 * sin(60)})(240:300:1.5)
\centerarc[](0,0)(0:360:.75)
\end{scope}
\end{tikzpicture}
\nonumber
\end{equation}
and so on for higher $n$. 

We now compute the banana diagrams for all $n\geq 2$, having already explicitly derived the one-loop result in Section~\ref{sec:oneloop} and the two-loop result in Section~\ref{sec:twoloop}. Importantly, the number of connecting lines between the two vertices is $n+1$ for the $n$-loop case. Each of these diagrams has either one internal propagator $G_{\varepsilon\varphi}$ or two. (Note that there is also one external propagator $G_{\varepsilon\varphi}$.) The two relevant types of integrals at each loop $n$ are therefore:
\begin{align}
I_n &= -n!\lambda_{n-1}^2 k^2 \int_{p_1,\dots,p_n} \,k_1^2 \ G_{\varepsilon\varphi}(p_1)G_{\varepsilon\varepsilon}(p_2) \nn
& \times G_{\varepsilon\varepsilon}(p_3)\dots G_{\varepsilon\varepsilon}(p_n) G_{\varepsilon\varepsilon}(p-p_1-\ldots-p_n)
\label{lambdasquaredBananaDiagram}
\end{align}
and
\begin{align}
\tilde{I}_n = & \ icT^2n^2(n-1)!\lambda_{n-1}\tilde\lambda_{n-1} k^2 \nn &\times\int_{p_1,\dots,p_n} \mathbf{k}_1 \cdot \mathbf{k}_2 \, G_{\varepsilon\varphi}(p_1) G_{\varepsilon\varphi}(p_2) \nn &\times G_{\varepsilon\varepsilon}(p_3)
\ldots G_{\varepsilon\varepsilon}(p_n) G_{\varepsilon\varepsilon}(p-p_1-\ldots-p_n),
\label{lambdalambdatildeBananaDiagram}
\end{align}
where we already took into account the symmetry factors of the two diagrams, which are $n!$ and $2(n-1)!$, respectively. We denote $\int_v \equiv \int\dots \frac{d^d v}{(2\pi)^d}$ for any $d$-dimensional vector $v$ of arbitrary signature.

To evaluate the $n$ frequency integrals over $\omega_1,\ldots,\omega_n$, we use Cauchy's integral formula and close the contour by a semicircular arc in the upper half-plane. The two types of integrals $I_n$ and $\tilde I_n$ then become a sum over the frequency-space poles of the integrand in the upper half-plane. In both cases, one obtains  the expression
\begin{align}
    \omega + iD\left(\left| \mathbf{k}-\sum_{i=1}^n\mathbf{k}_i\right|^2 + \sum_{i=1}^n k_i^2\right)
    \label{denominator result}
\end{align}
as the only $\omega$-dependent factor in the denominator. In other words, it is straightforward to check that one obtains \eqref{denominator result} by successive integrations over $\omega_i$, each time checking which poles are in the upper half-plane and using Cauchy's formula for each of those poles.

The integrals \eqref{lambdasquaredBananaDiagram} and \eqref{lambdalambdatildeBananaDiagram} then become
\begin{align}
I_n &= -n! \left(c T^2\right)^n \lambda_{n-1}^2 k^2 J_n\\
\tilde{I}_n &= \frac{n!\left(c T^2\right)^n}{n-1}\lambda_{n-1}\tilde\lambda_{n-1} k^2 \tilde{J}_n,
\end{align}
where
\begin{align}
J_n = \int_{\mathbf{k}_1,\dots,\mathbf{k}_n} \frac{\sum_{i=1}^nk_i^2}{\omega + iD\left(\left| \mathbf{k}-\sum_{i=1}^n\mathbf{k}_i\right|^2 + \sum_{i=1}^n k_i^2\right)}
\label{momentumIntegralJ1}
\end{align}
and
\begin{align}
\tilde{J}_n = \int_{\mathbf{k}_1,\dots,\mathbf{k}_n} \frac{\left|\sum_{i=1}^n{\bf k}_i\right|^2-\sum_{i=1}^n k_i^2}{\omega + iD\left(\left| \mathbf{k}-\sum_{i=1}^n\mathbf{k}_i\right|^2 + \sum_{i=1}^n k_i^2\right)}
\label{momentumIntegralJ2}
\end{align}
are the remaining momentum integrals on which we focus next. We note that in expressing them, we used a permutation symmetry among the momenta, replacing $k_1^2\rightarrow \frac{1}{n}\sum_{i=1}^n k_i^2$ in the numerator of \eqref{momentumIntegralJ1} and similarly for the integral $\tilde{J}_n$.

Focusing on the denominator structure \eqref{denominator result}, we now introduce a linear transformation mapping the spatial momenta $\mathbf{k}_1,\dots, \mathbf{k}_n$ to some $\mathbf{q}_1,\dots,\mathbf{q}_n$, such that
\begin{align}
    \mathbf{k}_i = a_{i0} \mathbf{k} + \sum_{j=1}^n a_{ij} \mathbf{q}_j,
    \label{q transform}
\end{align}
and use it in the two integrals above. To facilitate the spherical integration, we demand that all terms containing $\mathbf{k}\cdot\mathbf{q}_i$ in the denominator be zero and solve for the coefficients $a_{i0}$ and $a_{ij}$. One can show this results in
\begin{align}
    a_{10} = \ldots = a_{n0} = \frac{1}{n+1},
    \label{ai0 solution}
\end{align}
for the coefficients $a_{i0}$, and that
\begin{align}
    \left[\left(1-\sum_{i=1}^n a_{i0}\right)^2 + \sum_{i=1}^n a_{i0}^2\right] = \frac{1}{n+1}
\end{align}
is then the coefficient of $k^2$ in the denominator \eqref{denominator result}. Without discussing the solutions for the coefficients $a_{ij}$, we have sufficient information to show that such a transformation enables us to write the denominator \eqref{denominator result} in the following form:
\begin{align}
    \omega + \frac{i D k^2}{n+1} + f(\mathbf{q}_1,\dots,\mathbf{q}_n),
    \label{denominator result transformed}
\end{align}
where $f=f(\mathbf{q}_1,\dots,\mathbf{q}_n)$ is a function of the $\mathbf{q}_i$ alone. The two integrals in Eqs.~\eqref{momentumIntegralJ1} and \eqref{momentumIntegralJ2} therefore have the form
\begin{align}
     \int_{\mathbf{q}_1,\dots,\mathbf{q}_n} \frac{g(\mathbf{q}_1, \dots, \mathbf{q}_n)}{z_n^2 + f(\mathbf{q}_1, \dots, \mathbf{q}_n)},
     \label{full integral}
\end{align}
where we have introduced
\begin{equation}
    z_n^2 = k^2 - \frac{i(n+1)\omega}{D}.  
    \label{zn definition}
\end{equation}
It is now easy to see that the above expression \eqref{full integral}, which is a function of $z_n$ only, has a branch point at $z_n=0$. In other words, this proves that the $n$-loop contribution to the diffusive correlator (which always contains diagrams of the type discussed in this section) will have a branch point at the following location in the complex frequency plane:
\begin{align}
    \omega = -\frac{i D k^2}{n+1}.
\end{align}

This was a straightforward way to show that each loop correction introduces another branch point into the denominator of the retarded correlator and this statement is true regardless of the sizes of the $\lambda_i$ and $\tilde \lambda_i$ coefficients. Having shown the existence of the branch points, we now continue with the analysis of the banana diagrams in order to obtain explicit expressions containing the non-analyticities.

If we also demand that there are no mixed terms between different $\mathbf{q}_i$ in $f(\mathbf{q}_1, \dots, \mathbf{q}_n)$ of Eq.~\eqref{denominator result transformed}, then the coefficients $a_{ij}$ need to satisfy the equation
\begin{align}
    \sum_{i,i'=1}^n (1+\delta_{ii'})a_{ij}a_{i'j'}=0, ~\text{for all} ~ j\neq j',
\end{align}
where $\delta_{ii'}$ is the Kronecker delta. This equation can be solved by taking
\begin{align}
    a_{ij}=\frac{1}{\sqrt{n+1}}\left(\delta_{ij} - \frac{1}{n+1+\sqrt{n+1}}\right).
    \label{aij solution}
\end{align}
It follows that, with this choice of $a_{i0}$ and $a_{ij}$ (cf.~Eqs.~\eqref{ai0 solution} and \eqref{aij solution}), the momentum integral measure changes under the transformation \eqref{q transform} as
\begin{align}
    d^d{\bf k}_1\ldots d^d{\bf k}_n = (n+1)^{-\frac{(n+1)d}{2}}d^d{\bf q}_1\ldots d^d{\bf q}_n,
\end{align}
while the two momentum integrals become
\begin{align}
J_n &= -\frac{i(n+1)^{-\frac{(n+1)d}{2}}}{D} \nonumber \\
&\times \int_{\mathbf{q}_1,\dots,\mathbf{q}_n} \frac{\frac{n}{n+1}k^2 +\frac{n^2+n-1}{n(n+1)}\sum_{i=1}^n q_i^2 + R_n}{z_n^2+\sum_{i=1}^n q_i^2},
\end{align}
and
\begin{align}
\tilde{J}_n &=  -\frac{i(n+1)^{-\frac{(n+1)d}{2}}}{D} \nonumber \\
&\times \int_{\mathbf{q}_1,\dots,\mathbf{q}_n}\frac{\frac{n(n-1)}{n+1}k^2 - \frac{n-1}{n}\sum_{i=1}^n q_i^2 + \tilde{R}_n}{z_n^2+\sum_{i=1}^n q_i^2},
\end{align}
where $R_n$ and $\tilde{R}_n$ are polynomials in ${\bf k}$ and ${\bf q}_i$, which only comprise terms like ${\bf k}\cdot {\bf q}_i$ and ${\bf q}_i\cdot {\bf q}_j$, with $i\neq j$, and where we again used $z_n$ from Eq.~\eqref{zn definition}.

At this point, let us discuss the integration domain which has so far been suppressed. Initially, we thought of all integrals as having a hard momentum cut-off $\left|\mathbf{k}_i\right|<\Lambda$. This domain is transformed in a somewhat complicated way when we change the integration variables to ${\bf q}_i$, though, importantly, it still contains the origin since ${\bf k}$ is small compared to the cut-off. Since we are only interested in the non-analytic parts of the integrals as functions of $z_n$, we may replace the integration domain with any other domain containing the origin without affecting the non-analyticity, simply because the difference integral is obviously analytic around $z_n=0$ as it excludes a small ball around the origin.

We can thus perform another variable transform to the spherical coordinates of the full vector $({\bf q}_1,\ldots,{\bf q}_n)$ and take our integration domain to be an $nd$-dimensional ball of some radius $\tilde\Lambda$ around the origin. Note that now the $R_n$- and $\tilde{R}_n$-dependent terms disappear due to symmetry considerations. The full integration measure becomes
\begin{equation}
    \int_{\mathbf{q}_1,\dots,\mathbf{q}_n} \ \rightarrow \ \frac{\nu_{nd}}{(2\pi)^{nd}} \int r^{nd-1} dr,
\end{equation}
where we already took into account that the integrand does not depend on the angles in these new coordinates (due to the symmetric integration domain making $R_n$- and $\tilde{R}_n$-dependent terms zero) and $\nu_m$ is the volume of an $m$-sphere. We find\footnote{We continue using the equal sign in the below expressions even though those are no longer the full integrals. This is entirely sufficient for us to extract the cut-off-independent, non-analytic in $z$ parts of the integrals.}
\begin{align}
J_n &= -  \frac{i(n+1)^{-(n+1)d/2}\nu_{nd}}{(2\pi)^{nd}D}  \nonumber \\
&\times \int_0^{\tilde\Lambda} \frac{\frac{n}{n+1}k^2 + \frac{n^2+n-1}{n(n+1)}r^2}{z_n^2+r^2}r^{nd-1}dr,
\end{align}
and
\begin{align}
\tilde{J}_n &= -  \frac{i(n+1)^{-(n+1)d/2}\nu_{nd}}{(2\pi)^{nd}D} \nonumber \\
&\times \int_0^{\tilde\Lambda} \frac{\frac{n(n-1)}{n+1}k^2 - \frac{n-1}{n}r^2}{z_n^2+r^2}r^{nd-1}dr.
\end{align}

The non-analytic parts of these integrals now have a simple form when expressed in terms of 
\begin{align}
    \alpha_m(z) = \frac{(-1)^{\lfloor m/2\rfloor}}{(4\pi)^{m-1}} \frac{m}{4} z^{m-2} \cdot
    \begin{cases}
        1 & \text{$m$ odd}, \\
        \frac{1}{\pi} \log(z^2) & \text{$m$ even}.
    \end{cases}
\end{align}
The final expressions for the two integrals \eqref{lambdasquaredBananaDiagram} and \eqref{lambdalambdatildeBananaDiagram} are therefore
\begin{align}
I_n =& \, a_n \lambda_{n-1}^2 \alpha_{nd}(z_n) \nn 
&\times \left(\frac{n-1}{n(n+1)} k^4 - \frac{n^2+n-1}{n} \frac{i\omega k^2}{D} \right), \label{In_fin} \\
\tilde{I}_n =&\, a_n \lambda_{n-1}\tilde\lambda_{n-1}\alpha_{nd}(z_n) \nn
&\times\left(\frac{n^2+n+1}{n(n+1)} k^4 - \frac{n+1}{n} \frac{i\omega k^2}{D} \right), \label{Int_fin} 
\end{align}
where the prefactor $a_n$ is given by
\begin{equation}
    a_n = - \frac{in!(cT^2)^n(n+1)^{-\frac{(n+1)d}{2}}2^{nd}\nu_{nd}}{2Dnd}.
\end{equation}
The above expressions hold for all $n\geq2$, whereas the one-loop results were calculated in Ref.~\cite{Chen-Lin:2018kfl} and in Appendix \ref{app:details1loop}. For the sake of completeness, we reiterate them here:
\begin{align}
    I_1 &= -\frac{cT^2\nu_d}{2D} \lambda_0^2 k^2 \omega \alpha_d(z_1) ,\label{I1_fin}\\
    \tilde{I}_1 &= -\frac{cT^2\nu_d}{2D} \lambda_0 \tilde\lambda_0 \left(w+iDk^2\right)\alpha_d(z_1). \label{I1t_fin}
\end{align}
Finally, we note again that it is important to keep in mind that beyond these non-analyticities that enter into the denominator of diffusive $G_R$, the diffusivity $D$ also receives an infinite number of cut-off dependent corrections.

\section{Analysis of the $n$-loop banana diffusion results and discussion}
\label{sec:nloopbranchcutAnalysis}

In the final section of this paper, we briefly review the results, discuss a number of simple implications and observations related in particular to the $n$-loop results derived in Section~\ref{sec:nloopbranchcut} and finish by mentioning a few open problems. 

In this work, we considered the structure of retarded diffusive hydrodynamic correlators $G_R$ as computed to all orders in the loop expansion from the first-order (in the gradient expansion) effective hydrodynamic theory. The general picture that has emerged in this work and in previous works (see Refs.~\cite{Chen-Lin:2018kfl,Delacretaz:2020nit,Michailidis:2023mkd}) is that the classical (tree-level) diffusive pole tends to split into two poles, although the reliability of this phenomenon, which depends on powers of $k$, depends on the dimensionality of space. In particular, the correction splitting the tree-level pole with the diffusive dispersion relation $\omega \sim k^2$ is proportional to $\pm k^{2+nd}$, where $n$ is the order of the loop expansion and $d$ the number of spatial dimensions. Since the classical (tree-level) gradient expansion itself produces corrections to $\omega(k)$ that scale with $k^4$, the only correction that can dominate over the $k^4$ terms exists for $d=1$ and $n=1$. Beyond this term, we cannot at this point unequivocally establish the `splitting' of the diffusive pole as this would require the inclusion of higher-derivative corrections to the effective action. 

In general, we found that, due to the structure of the self-energy $\Sigma$, the diffusive dispersion relation combines an analytic and a non-analytic piece encoded in the $i\delta Dk^2$ and $\Sigma_*$ terms (cf.~Eq.~\eqref{sigma1loop}). Importantly, the analytic pieces are cut-off dependent and renormalise the diffusion constant. On the other hand, the loop-corrections to $\Sigma_*$ include cut-off independent terms and qualitatively change the analytic structure of $G_R$ by introducing an infinite number of branch points \eqref{BranchPoint}. We note that even though we did not explicitly compute the non-analytic structure of the numerator (i.e., of $\chi_*$), we do not expect those corrections to introduce any new types of non-analyticities into $G_R$

Then, to keep the all-loop analysis as concrete as possible, we focused only on the non-analytic cut-off-independent terms with a specific banana diagram scaling of coupling constants. In this limit, the leading-order corrections to the tree-level diffusion are all represented by the $n$-loop banana diagrams. We found that in any theory of diffusion appropriately described by such a limit, the denominator of $G_R$ has the following form:
\begin{align}
    \omega+i D_R k^2 + \sum_{n=1}^\infty \left(I_n + \tilde{I}_n\right),
    \label{full denominator}
\end{align}
where $D_R$ is the `renormalised' and cut-off-dependent diffusion constant,
while all the higher-loop non-analytic corrections to the self-energy are contained in the infinite sum. The closed-form results of $I_n$ and $\tilde I_n$ are stated in Eqs.~\eqref{I1_fin} and \eqref{I1t_fin} for $n=1$ and in Eqs.~\eqref{In_fin} and \eqref{Int_fin} for $n\geq 2$.
  
When adding the higher-loop corrections, one immediately notices that the analytic structure of the denominator quickly becomes quite complicated. With each summand in the infinite sum of \eqref{full denominator}, the denominator obtains a new branch point, which  is either algebraic (due to the square root) or logarithmic (due to the logarithm). The first case introduces two new branches in the Riemann surface, while the logarithmic branch point results in a countably infinite set of branches at a specific $z_n$ branch point.

\begin{figure}[ht!]
\centering
\includegraphics[width=\linewidth]{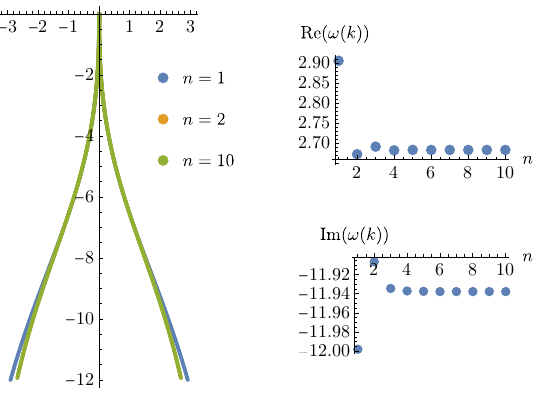}
\caption{\textbf{Left:} Hydrodynamic dispersion relation $\omega(k)$ for $k\in[0,4]$ in $d=3$ plotted in the complex $\omega$-plane for results up to $n=1$, $2$ and $10$ loops, with all $\gamma_{1\leq i\leq n}$ negative (we set all parameters equal to one). Due to fast convergence, the results to $n=2$ and $n=10$ orders are indistinguishable on this plot. \textbf{Right:} Real and imaginary parts of the dispersion relation plotted for different truncations of $n$-loop contributions for $k=4$ again showing rapid convergence.}
\label{fig:disprel}
\end{figure}

Next, we consider in more detail the hydrodynamic dispersion relations expanded in small $k$. We find the all-loop series of corrections to the diffusive tree-level pole $\omega = -iD_R k^2 \pm \delta\omega$ to have the following structure (cf.~Eq.~\eqref{smallwkpoleassumption}):
\begin{align}
    \delta \omega = \sum_{n=1}^\infty c_n\gamma_n^{\frac{nd}{2}-1} k^{2+nd} \begin{cases} 
        1, & n d\text{ odd}, \\
        \log(\gamma_n k^2), & n d\text{ even},
    \end{cases}
    \label{delta omega all loop}
\end{align}
where $c_n$ are easily-computable coefficients that we do not state explicitly and where
\begin{equation}
    \gamma_n = 1-\frac{(n+1)D_R}{D}.
\end{equation}
This result therefore generalises the one-loop (pole-splitting) result from Eq.~\eqref{delta omega 1 loop}. We show explicitly an example of this pole splitting that occurs in the presence of higher-loop corrections due to the structure of Eq.~\eqref{delta omega all loop} in Figure~\ref{fig:disprel}. It is qualitatively analogous to the one observed at one loop. 

\begin{figure*}[ht!]
\centering
\includegraphics[width=\linewidth]{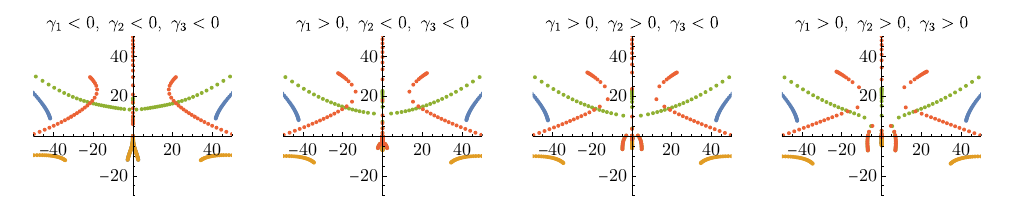}
\caption{Poles of the retarded correlator in the complex $\omega$-plane computed to $n=3$ loops in all four qualitatively distinct parameter cases for real $k\in[0,4]$. The value of $D'$ is $1$, $5/12$, $7/24$ and $1/8$, going from left to right, while all other parameters are set to one. One can notice the splitting of the pole from Figure~\ref{fig:disprel} (first found in Ref.~\cite{Chen-Lin:2018kfl}) on the leftmost plot in orange.}
\label{fig:n3d3}
\end{figure*}

Next, we note that at $n$ loops, many different choices for the sign of $\gamma_i$ exist. Now, each $\gamma_i$ for $i=1,\dots,n$ can be either positive or negative, with the condition that forbids any $\gamma_i > 0$ and simultaneously any $\gamma_j < 0$ for $j<i$. We should therefore consider $n+1$ different cases ranging from all $\gamma_i < 0$, to $\gamma_1 > 0$ positive and $\gamma_{i>1} < 0$, $\gamma_1 > 0$ and $\gamma_2 > 0$ and $\gamma_{i>2} < 0 $ negative, etc., up to all $\gamma_i > 0$. An example of such different choices for a three-loop result, again, treating the analytic structure in the complex $\omega$-plane non-perturbatively in $k$, is shown in Figure~\ref{fig:n3d3}.

\begin{figure}
\centering
\includegraphics[width=0.8\linewidth]{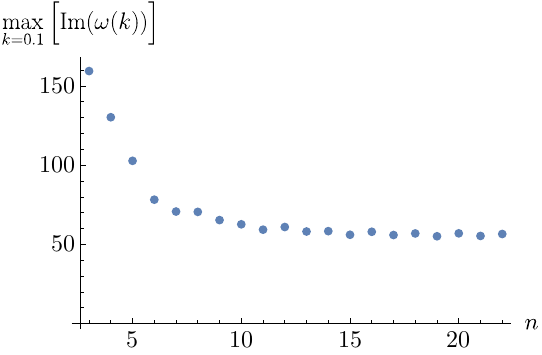}
\caption{Largest positive imaginary frequency among all the modes in the spectrum for $k=0.1$ plotted as a function of $n$ (the number of loops included in the non-analytic part of $\Sigma$). As in Figure~\ref{fig:n3d3}, we only vary $D'$ from $1$ going over all arithmetic means between all existing branch points to $1/2(n+1)$ in the $n$-loop case. All other parameters are set to one.}
\label{fig:MaxIm n dependence}
\end{figure}

In a further `non-perturbative exercise' using the full analytic structure of $G_R$, which may be of relevance to potential future discussions of non-perturbative stability of the hydrodynamic EFT,
we point out that there exist poles with dispersion relations that have $\im \,\omega > 0$ for real $k$. Note that this is already seen in Figure~\ref{fig:n3d3}. As is usual in a linear analysis of spectra, such modes signal an instability. Its strength can be quantified by plotting the largest $\im \,\omega > 0$ of any of the modes in the spectrum. We shown an example of its behaviour for truncations of the spectrum at different orders of loops in Figure~\ref{fig:MaxIm n dependence}.

Having analysed the non-analytic contributions to the self-energy to all orders in $n$, several outstanding questions for the full, physical analysis of diffusion remain. The computation of the remaining diagrams in \eqref{nonFactorizableTwoLoop} to $n$-loop order is still outstanding, although we expect that the analytic structure arising from such terms would not dramatically modify $G_R$ computed in the banana diagram limit. Of particular future interest should then be potential resummations of various diagrams, even just those computed explicitly in this work. Another natural direction of research is the analysis of a CTP EFT of sound. Moreover, it would also be intriguing to see how the effective diffusive action \eqref{Lint} interacts with additional non-hydrodynamic modes, in particular modes which become relevant near the boundary of a phase transition. This would nicely complement recent works using stochastic hydrodynamics near the chiral phase transition \cite{Grossi:2020ezz,Grossi:2021gqi}. Finally, another interesting direction of research may be the analysis of loop corrections arising from actions that go beyond the (classical) structures studied here, in particular, by including effects discussed in Ref.~\cite{Jain:2020zhu}.

\begin{acknowledgments}
We would like to thank Luca Delacr\'{e}taz, Sean Hartnoll and Pavel Kovtun for illuminating discussions. The work of S.G. was supported by the STFC Ernest Rutherford Fellowship ST/T00388X/1. The work is also supported by the research programme P1-0402 and the project N1-0245 of Slovenian Research Agency (ARIS). T.L. is supported by the research programme P1-0402 of Slovenian Research Agency (ARIS). J.P. was supported by the Clarendon Fund scholarship. A.S. was supported by funding from Horizon Europe research and innovation programme under the Marie Sk\l{}odowska-Curie grant agreement No. 101103006% and the project N1-0245 of Slovenian Research Agency (ARIS)
.
\end{acknowledgments}

\appendix

\section{Details of the one-loop calculation}
\label{app:details1loop}

We first write down the full expressions for the two functions $C$ and $\Sigma$ that we introduced in \eqref{GsymExpression} for the symmetrised energy density correlator $G_S$. 

For the numerator $C$ and to one loop,
\begin{align}
    &C(\omega, k) = 2\kappa T^2 k^2 + 2cT^2k^2 \tilde\lambda' \int_{p'} G_{\varepsilon\varepsilon}(p')  \nonumber \\
    &- \frac{1}{2}k^4 \lambda^2 \int_{p'} G_{\varepsilon\varepsilon}(p') G_{\varepsilon\varepsilon}(p-p') \nonumber \\
    &+ \left(2icT^2k^2 \lambda \tilde\lambda \int_{p'} \mathbf{k}\cdot\mathbf{k}' G_{\varepsilon\varphi}(p') G_{\varepsilon\varepsilon} (p-p') + \text{c.c.} \right),
    \label{C_integrals}
\end{align}
where the first term is the tree-level result.

For the function $\Sigma$, we have
\begin{align}
    &\Sigma(\omega, k) = i k^2 \lambda' \int_{p'} G_{\varepsilon\varepsilon}(p') \nn
    &- k^2 \lambda^2 \int_{p'} k'^2 G_{\varepsilon\varphi}(p') G_{\varepsilon\varepsilon} (p-p') \nn
    &+ icT^2k^2 \lambda \tilde\lambda \int_{p'} \mathbf{k}'\cdot (\mathbf{k} - \mathbf{k}') G_{\varepsilon\varphi}(p') G_{\varepsilon\varphi} (p-p').
\end{align}
Note that we employed the following shorthand notation for the integrals:
\begin{equation}
    \int_{p'} = \int \frac{d\omega}{2\pi} \frac{d^dk}{(2\pi)^d}.
\end{equation}

The integrals over $\omega$ are straightforward and can be done using Cauchy's formula. The subsequent momentum integrals are evaluated in spherical coordinates and new variables $\mathbf{q}$ and $z$, where
\begin{align}
    \mathbf{k}' &= \frac{1}{2}\left( \mathbf{k} + \mathbf{q} \right), \\
    z^2 &= k^2 - \frac{2i\omega}{D} \label{zVariable}.
\end{align}
Then, the one-loop integrals can all be written in terms of a single family of integrals indexed by $n\in\mathbb{N}$ with one parameter $z$,
\begin{align}
    \mathcal{I}_n(z) = \int_0^\Lambda \frac{q^n}{q^2 + z^2} dq.
\end{align}
All of these integrals are in principle multivalued functions. We do not want our correlators to be multivalued, therefore we have to decide which contour in the complex $q$-plane to integrate over. In this way, we determine the branch and the branch cut we use for the correction to the correlator. We choose the principal value prescription, which also coincides with the choice that results in all of the poles being on the same branch (see Ref.~\cite{Bajec:2024jez}). 

Keeping only the highest order terms in $\Lambda$ in both the analytic and non-analytic part, we get
\begin{align}
    \mathcal{I}_n = \frac{\Lambda^n}{n-1} +  \frac{(-1)^{\lceil n/2\rceil}\pi}{2} z^n \begin{cases}
        1, & n \text{ even}, \\
        \frac{1}{\pi}\log(z^2), & n \text{ odd}.
    \end{cases}
\end{align}

Using the above results, we find
\begin{align}
    \Sigma = i \delta D k^2 + \Sigma_{\star},
\end{align}
where the cut-off dependent term $i\delta D k^2$ corrects the diffusion constant, while the cut-off independent non-analytic term $\Sigma_\star$ introduces a new branch point into the denominator of the propagator.

The new diffusion constant $D'$ is then
\begin{align}
    D' = D + \delta D,
\end{align}
where
\begin{align}
    \delta D = \frac{cT^2\nu_d}{2D(2\pi)^d} \left(\lambda(\lambda+\tilde\lambda) - 2\lambda' D\right)\Lambda^d.
\end{align}
The non-analytic correction to the denominator is
\begin{align}
    \Sigma_\star = -\frac{cT^2\nu_d}{2D^2} k^2 \left(\omega\lambda(\lambda+\tilde\lambda) + iDk^2 \lambda\tilde\lambda\right) \alpha_d,
    \label{sigmastar1loop}
\end{align}
where
\begin{align}
    \alpha_d(z) = \frac{(-1)^{\lfloor d/2\rfloor}}{(4\pi)^{d-1}} \frac{d}{4} z^{d-2} \cdot
    \begin{cases}
        1 & \text{$d$ odd}, \\
        \frac{1}{\pi} \log(z^2) & \text{$d$ even}.
    \end{cases}
\end{align}

One can now determine the position of the two poles to first order by using the ansatz $\omega = -iD'k^2 \pm \delta\omega$ and by solving the equation
\begin{align}
    \pm\delta\omega + \Sigma_\star \left( -iD'k^2\pm\delta\omega, k\right) = 0.
\end{align}
We obtain
\begin{align}
    \delta \omega = \frac{icT^2\nu_d}{2D^2} k^{d+2} \gamma^{d/2 - 1} \left( D'\lambda(\lambda + \tilde\lambda) - D \lambda\tilde\lambda \right) \beta_d,
\end{align}
where
\begin{align}
    \beta_d(k) = \frac{(-1)^{\lfloor d/2\rfloor}}{(4\pi)^{d-1}}\frac{d}{4} \cdot
    \begin{cases}
        1 & d \text{ odd}, \\
        \frac{1}{\pi} \log(\gamma k^2) & d \text{ even},
    \end{cases}
\end{align}
and
\begin{align}
    \gamma = 1-\frac{2D'}{D}.
    \label{gammaExpression}
\end{align}
The exact manner in which the tree-level pole $\omega=-iDk^2$ splits depends on the parameter values. For $d$ odd, $\delta\omega$ is real if $\gamma>0$ and purely imaginary if $\gamma<0$. For $d$ even, $\delta\omega$ is complex if $\gamma>0$ and again purely imaginary if $\gamma<0$.

\section{One-loop corrections for a gapped theory}\label{app:gapped}

It is instructive (and straightforward) to repeat the above calculation for the case of a gapped dispersion relation,
\begin{equation}
\omega = -ia_0 - iDk^2.    
\end{equation}
Omitting some steps in the calculation, we take the free propagators in \eqref{props} to have the form
\begin{align}
\gef(p) &= \frac{1}{\omega + i(a_0 + Dk^2)},\nonumber\\
\gfe(p) &= \frac{1}{-\omega + i(a_0 + Dk^2)},\nonumber\\
\gee(p) &= \frac{2cT^2 (a_0 + Dk^2)}{\omega^2 + (a_0 + Dk^2)^2},\nonumber\\
G_{\varphi\varphi}(p) &= 0\label{gappedProps}
\end{align}
and repeat the analysis from Section~\ref{sec:oneloop}.

Evaluating the integrals is done in exactly the same manner as in \ref{app:details1loop}, except that the variable $z$ in \eqref{zVariable} is in this case taken to be
\begin{equation}
    z^2 = k^2 - \frac{2i\omega}{D} + \frac{4a_0}{D}\label{newzVariable}.
\end{equation}

The function $\Sigma$ is again of the same form,
\begin{align}
    \Sigma = i \delta D k^2 + \Sigma_{\star},
\end{align}
where $\delta D$ is the same as in the ungapped case, while the cut-off independent non-analytic term becomes
\begin{align}
    \Sigma_\star = -\frac{cT^2\nu_d}{2D^2} k^2 \left((\omega+2ia_0)\lambda(\lambda+\tilde\lambda) + iDk^2 \lambda\tilde\lambda\right) \alpha_d,
\end{align}
where
\begin{align}
    \alpha_d(z) = \frac{(-1)^{\lfloor d/2\rfloor}}{(4\pi)^{d-1}} \frac{d}{4} z^{d-2} \cdot
    \begin{cases}
        1 & \text{$d$ odd}, \\
        \frac{1}{\pi} \log(z^2) & \text{$d$ even},
    \end{cases}
\end{align}
with the new $z$ from \eqref{newzVariable}.

One can now again determine the position of the two poles to first order by using the ansatz $\omega = -iD'k^2 \pm \delta\omega$ and by solving the equation
\begin{align}
    \pm\delta\omega + \Sigma_\star \left( -iD'k^2\pm\delta\omega, k\right) = 0.
\end{align}
In this case, we obtain
\begin{align}
    \delta \omega &= \frac{ick^2T^2\lambda \nu_d}{2D^2} \left( \gamma k^2 + \frac{4a_0}{D}\right)^{d/2-1} \nn
    &\times \left( (\lambda + \tilde\lambda)(D'k^2-2a_0) - D k^2 \tilde\lambda \right) \beta_d, 
\end{align}
where
\begin{align}
    \beta_d(k) = \frac{(-1)^{\lfloor d/2\rfloor}}{(4\pi)^{d-1}}\frac{d}{4} \cdot
    \begin{cases}
        1 & d \text{ odd}, \\
        \frac{1}{\pi} \log\left(\gamma k^2 + \frac{4a_0}{D}\right) & d \text{ even},
    \end{cases}
\end{align}
with the factor $\gamma$ being the same as in \eqref{gammaExpression}.

\bibliography{bib}

\end{document}